\documentclass[onecolumn]{article}
\usepackage{authblk}
\usepackage[a4paper, total={6in, 8in}]{geometry}

\usepackage[
backend=biber,
style=numeric,
sorting=none
]{biblatex}
\usepackage{graphicx} 
\usepackage{xcolor}
\usepackage{subcaption}
\usepackage{booktabs}

\usepackage{setspace}
\DeclareUnicodeCharacter{200B}{}

\title{Impact of model uncertainty on SPARC operating scenario predictions with empirical modeling}

\author[1]{A. Saltzman}
\author[1]{P. Rodriguez-Fernandez}
\author[2]{T. Body}
\author[1]{A. Ho}
\author[1]{N.T. Howard}
\affil[1]{MIT Plasma Science and Fusion Center}
\affil[2]{Commonwealth Fusion Systems}
\date{}

\begin{document}

\maketitle

\begin{abstract}

Understanding and accounting for uncertainty helps to ensure next-step tokamaks such as SPARC will robustly achieve their goals. While traditional Plasma OPerating CONtour (POPCON) analyses guide design, they often overlook the significant impact of uncertainties in scaling laws, plasma profiles, and impurity concentrations on performance predictions. This work confronts these challenges by introducing statistical POPCONs, which leverage Monte Carlo analysis to quantify the sensitivity of SPARC's operating points \cite{creely_overview_2020} to these crucial variables. For profiles, a physically motivated gradient-based functional form is introduced. We further develop a multi-fidelity Bayesian optimization workflow that effectively identifies operating points maximizing the \textit{probability} of meeting performance goals, which gives a significant speed-up over brute force methods. Our findings reveal that accounting for these uncertainties leads to an optimal operating point different from deterministic predictions, which balances H-mode access, confinement, impurity dilution, and auxiliary power. 
\end{abstract}

\section{Introduction}

New tokamak experiments and power plants have historically been designed starting from empirically-based Plasma OPerating CONtour (POPCON)-like analyses \cite{creely_overview_2020, houlberg_fusion_1982, doyle_chapter_2007,han_start-up_2013,  sorbom_arc_2015, frank_radiative_2022, manta_collaboration_manta_2024}. POPCONs are a tool to predict how tokamak performance changes when moving around the operating space. Traditionally, this operating space is either volume-averaged, or on-axis, density and temperature. POPCONs allow rapid  inspection of boundaries including confinement mode transitions, disruption limits, input power limits, and radiative power limits to see if there is a stable scenario with the desired fusion power and fusion gain. 

To determine fusion power and fusion gain for each operating point, POPCONs solve for power balance using empirical scaling laws to predict energy confinement time \cite{iter_physics_expert_group_on_confinement_and_transport_chapter_1999, yushmanov_scalings_1990, kaye_iter_1997} and confinement transition thresholds \cite{martin_power_2008, ryter_experimental_2014}. However, there are large root mean square errors on these scaling laws (even when evaluated on machines for which these scalings have been developed) and a scalar multiple (\textit{H}) is often used to make adjustments to reflect performance variations among different confinement regimes. Assumed degradations or improvements in energy confinement time, which take the form of these scalar multiple ad-hoc factors, have dramatic implications for fusion device performance and are generally poorly constrained.  

In addition to assuming ad-hoc factors, POPCON workflows require assumptions on the profile forms of both plasma density and temperature in order to calculate volume integrate quantities, like total fusion power. Traditionally, parabolic profiles are assumed. However, these do not capture the convex behavior of the profile in the outer core, seen especially in H-modes, for experimental profiles and physics-based profile predictions. 

In this work, we explore how accounting for uncertainty can highlight operating scenarios that have not just high performance, but also a high confidence of high performance. We perform this analysis for SPARC, a tokamak under construction by Commonwealth Fusion Systems in Devens Massachusetts \cite{creely_overview_2020}. It has a major radius of 1.85 m. The baseline operating scenario for SPARC is an H-mode with a magnetic field strength of 12.2 T and a 8.7 MA plasma current \cite{rodriguez-fernandez_overview_2022}. The optimal operating points found in this work account for both the propagation of input parameter uncertainties and a redefinition of the optimization metric from maximizing fusion power within the operational window to maximizing pointwise success rate.  Section \ref{sec:Monte_Carlo_Methods} discusses uncertainty propagation through POPCON predictions via Monte Carlo analysis. Section \ref{sec:Statistical_POPCON_Results} presents statistical POPCONs using nominal assumptions for the SPARC Primary Reference Discharge (PRD). Section \ref{sec:Bayesian_Optimization_Methods} introduces the Bayesian optimization methods used to facilitate scans of assumptions. Section \ref{sec:Scanning_Results} presents the results of these scans, and Section \ref{sec:Discussion} provides a discussion of the implications for SPARC's operational planning.

\section{Uncertainty propagation in POPCONs} \label{sec:Monte_Carlo_Methods}

\begin{figure}[h]
    \centering
    \includegraphics[width=0.5\linewidth]{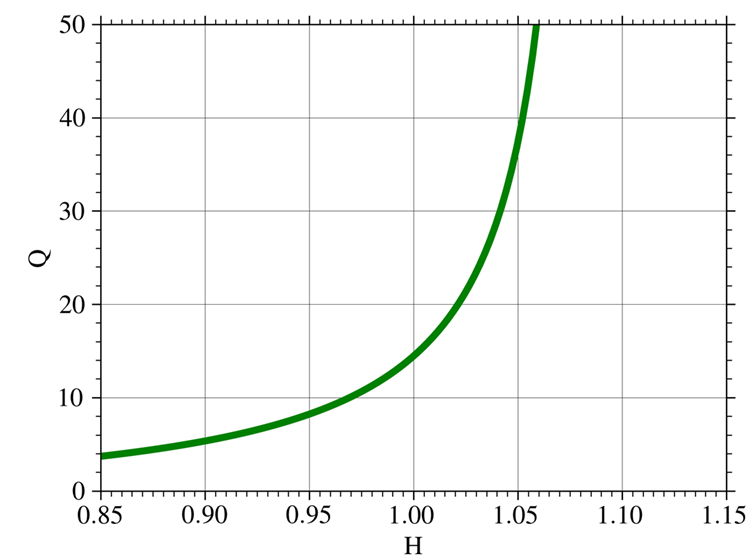}
    \caption{Dependence of fusion gain ($Q$) on energy confinement scalar multiple ($H$) of ITER98(y2) H-mode scaling law \cite{iter_physics_expert_group_on_confinement_and_transport_chapter_1999} for the SPARC PRD \cite{creely_overview_2020}. A strong sensitivity is observed, even with small deviations from H=1.}
    \label{fig:Q_vs_H}
\end{figure}

As a burning plasma device, SPARC is especially sensitive to confinement regime and profile assumptions. This has been shown in the change in predicted performance as modeling has increased fidelity from the original POPCON predictions \cite{creely_overview_2020}, to medium fidelity predictions \cite{rodriguez-fernandez_predictions_2020}, and finally to high-fidelity predictions \cite{rodriguez-fernandez_nonlinear_2022}. As shown in Figure \ref{fig:Q_vs_H}, a 15\% deviation, in line with the reported error on the ITER98(y2) H-mode energy confinement scaling \cite{iter_physics_expert_group_on_confinement_and_transport_chapter_1999}, in the energy confinement time for the SPARC PRD, results in a non-linear change in fusion gain from four to ignition. This motivates the need for careful uncertainty propagation.

\begin{figure}[h]
    \centering
    \includegraphics[width = \linewidth]{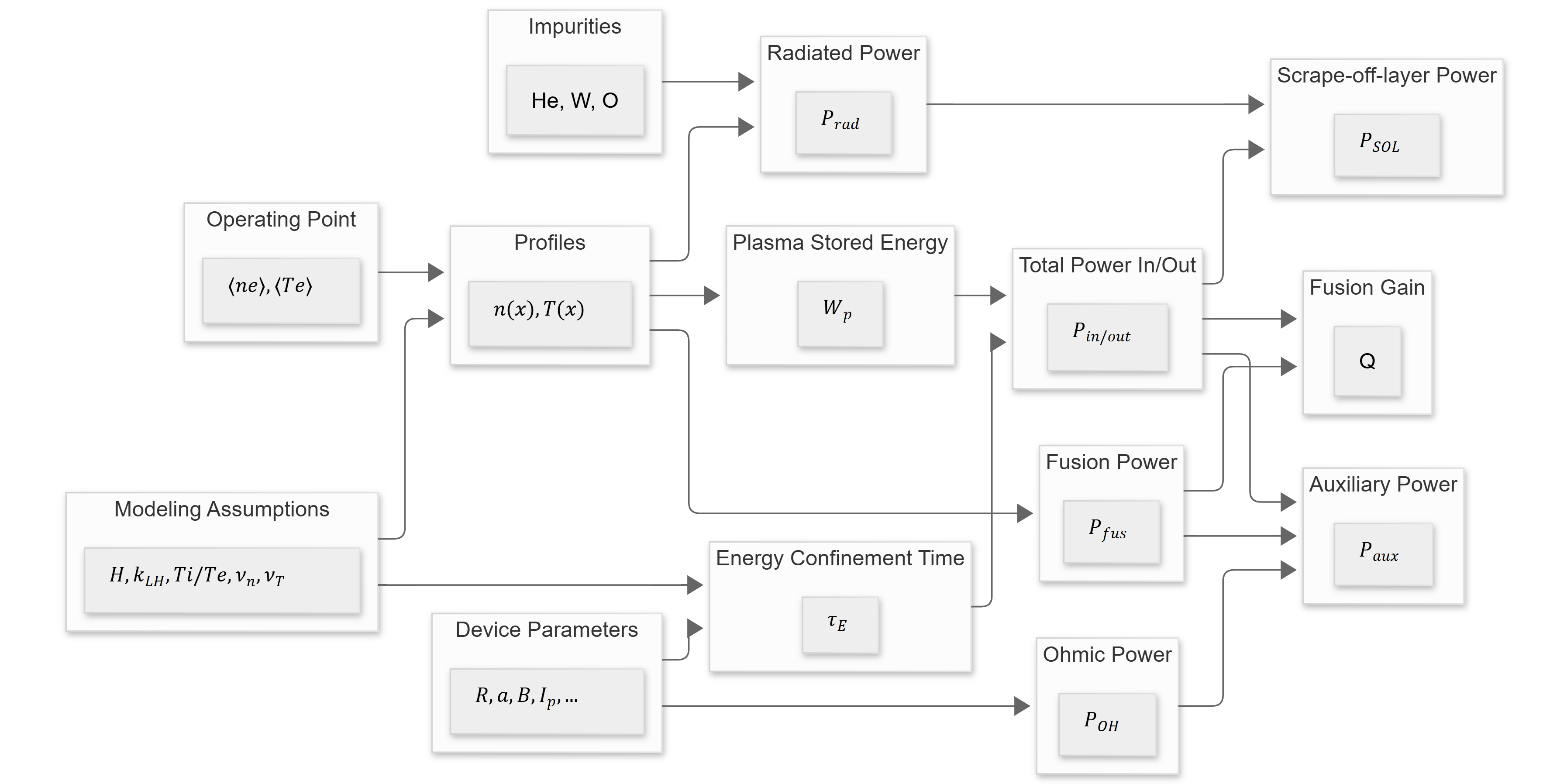}
    \caption{POPCONs work by solving a power balance workflow for a given operating point in, typically, volume-averaged density, volume-averaged temperature space. They require the device engineering parameters as well as profile and confinement assumptions to be input. This figure shows one possible POPCON workflow, representative of that traditionally used in CFSPOPCON \cite{body_cfs-energycfspopcon_2024}.} 
    \label{fig:POPCON_workflow}
\end{figure}

POPCON predictions are made in this work using CFSPOPCON  \cite{body_cfs-energycfspopcon_2024}, an open source POPCON tool. The general POPCON modeling workflow is shown in Figure \ref{fig:POPCON_workflow}. At each operation point in volume-averaged density, volume-averaged temperature space, the power balance workflow is independently solved. From the volume-averaged values and assumptions on shape, plasma profiles are generated in the native coordinate system of the POPCON code. For a further discussion of how we generate plasma profiles in this work, please see Section \ref{sub_sec:PRF_functionals}. From the calculated profiles, the fusion power and plasma stored energy are determined.

Engineering parameters including plasma geometry, magnetic field strength, and plasma current must be entered into the POPCON workflow. These in turn are input into empirical scaling laws to calculate quantities such as the energy confinement time ($\tau_E$). Combining knowledge of total plasma stored energy ($W_p$) with the energy confinement time ($\tau_E$) allows for the determination of the total power input into the plasma ($P_{in/out}$) according to $P_{in/out} = \frac{W_p}{\tau_E}$.  Hence, we can now solve the rest of the power balance equation $P_{in/out} = P_{OH} + P_\alpha + P_{aux}$, where $P_{OH}$ represents ohmic power, $P_\alpha$ represents the fusion power deposited in the alpha particles. Finally, the necessary quantity of auxiliary power ($P_{aux}$) coupled into the plasma and the fusion gain ($Q$) are determined.

The radiated power is determined based on the plasma impurities, which can be specified in terms of concentrations relative to the electron density, effective ion charge state ($Z_{eff}$), and dilution, or a fixed fraction of the L-H transition threshold power. The order of the boxes on the bottom row of Figure \ref{fig:POPCON_workflow} can be flipped depending on your chosen assumption. Since we are considering the steady state condition of tokamaks, power input into the plasma must be the same as the power which comes out of the plasma. Therefore, from the radiated power ($P_{rad}$) and total input/output power ($P_{in/out}$), we can determined the scrape-off-layer power ($P_{SOL}$) using $P_{in/out} = P_{rad} + P_{sol}$. A common error in POPCON analyses is enforcing that, at low temperatures, the radiative power is at most the total output power. In fact, if the calculated radiative power ($P_{rad}$) is greater than the total output power ($P_{in}$), the plasma is in radiative collapse and there is no steady state solution. Some amount of power will always be conducted out through the plasma by neoclassical and turbulent transport processes.

Following previous empirical modeling studies of SPARC H-modes \cite{body_cfs-energycfspopcon_2024, body_sparc_2023}, this work also utilizes fixed concentrations of tungsten, helium-3, and oxygen impurities. In addition, argon is introduced to help mitigate divertor heat fluxes. The amount of argon is calculated self-consistently with the boundary solution, which uses the Lengyel-Goedheer model \cite{stangeby_basic_2018} with a target electron temperature of 25 eV and an edge-core argon concentration enrichment factor of five \cite{kallenbach_divertor_2024}. Since impurity concentrations are an input, the L-H fraction is an output of the model. 

The confinement scaling law used to predict energy confinement time is chosen for each operating point based on whether the scrape-off-layer (SOL) power is above or below the L-H transition threshold. The L-H transition threshold was calculated via the Martin scaling, equation 2 in \cite{martin_power_2008} with the Ryter density minimum \cite{ryter_experimental_2014} and a penalty for the low density branch as in \cite{hughes_projections_2020}. A $2/A$, where A is the main ion mass number, was also applied per \cite{righi_isotope_1999}. The H-mode ITER98(y2) scaling law \cite{iter_physics_expert_group_on_confinement_and_transport_chapter_1999} was used for plasmas with a scrape-off-layer power greater than the L-H power threshold, and the L-mode ITER89p scaling law \cite{yushmanov_scalings_1990} was used for those with lower power. Additionally, since the fusion power in SPARC is constrained to be less than 140 MW, for a 10 second discharge, by magnet engineering considerations \cite{creely_overview_2020, rodriguez-fernandez_predictions_2020}. We call the POPCON diagrams with error propagated through ``statistical POPCONs".

\subsection{Physics-guided profile parameterization}\label{sub_sec:PRF_functionals}

Overly simplistic functional forms for the kinetic profiles ---such as  the parabolic shapes typically used for POPCON-type analyses--- can lead to unrealistic representations of core plasmas \cite{rodriguez-fernandez_enhancing_2024}. This is particularly important for H-mode operation and in burning plasma regimes, as unrealistic edge conditions or core gradients might result from the choice of volume-averaged temperatures. As an example, for the SPARC PRD, empirical modeling using parabolic profiles \cite{creely_overview_2020} and physics-based simulations with TRANSP \cite{rodriguez-fernandez_predictions_2020} showed non-negligible differences in fusion power ($\sim20\%$) even if global quantities were almost a perfect match. 

\begin{figure}[h]
    \centering
    \begin{subfigure}[t]{0.49\linewidth}
        \centering
        \includegraphics[width=\linewidth]{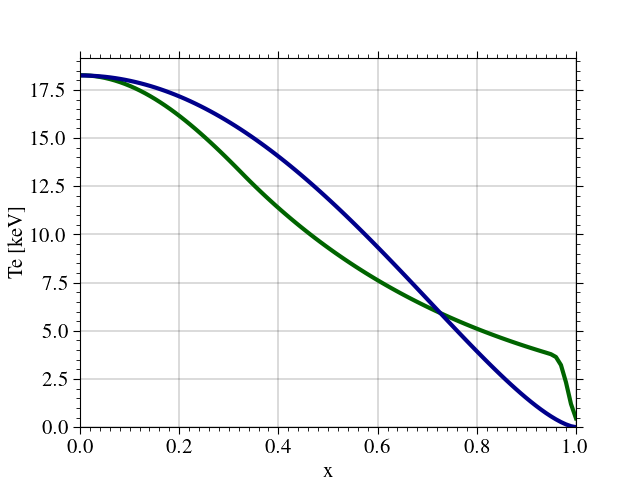}
        \caption{}
        \label{subfig:profile_PRF_functionals}

    \end{subfigure}
    \hfill
    \begin{subfigure}[t]{0.49\linewidth}
        \centering
        \includegraphics[width=\linewidth]{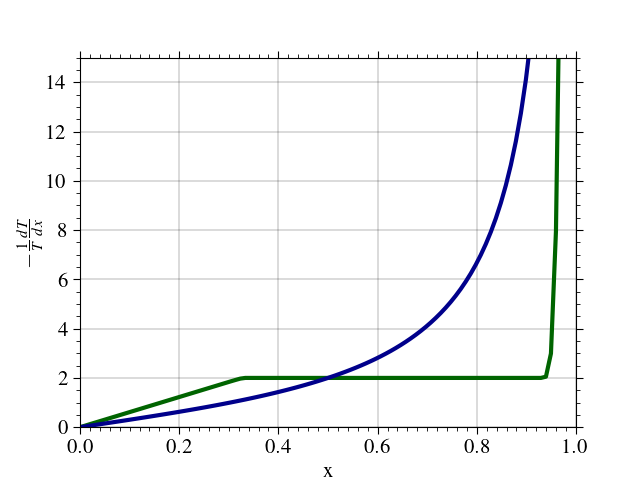}
        \caption{}
        \label{subfig:grad_PRF_functionals}
    \end{subfigure}
    \hfill
        \caption{For the same volume-averaged temperature and peaking, (a) profile and (b) inverse gradient scale lengths are plotted for (blue) parabolic and (green) gradient based functional forms.}
        \label{fig:PRF_functionals}
\end{figure}

In this work, we decided to use more physically motivated profile forms called gradient based functional forms, an example of which can be seen in Figure \ref{fig:PRF_functionals}. To be accommodated within POPCON-type workflows, these functional forms were designed to take similar input parameters: temperature peaking ($\nu_T$), density peaking ($\nu_{n}$), ion-to-electron temperature ratio ($T_i/T_e$), volume-averaged temperature ($\left<T\right>$), and volume-averaged density ($\left<n_e\right>$). However, a new parameter is included: the core inverse normalized temperature gradient scale length represented as a single scalar, $a/L_T$. The ion and electron temperature profiles are identical up to the $T_i/T_e$ scalar. 

To generate such profiles, it is assumed that the pedestal has a width of 5\% of the minor radius, such that $x_{ped}=0.95$. It is then assumed that the core inverse normalized temperature gradient scale length profile grows linearly from $x=0$ to $x=x_{inf}$, where it takes the constant value of the core inverse normalized temperature gradient scale length, $a/L_T$, up to $x_{ped}$. $x_{inf}$ is the profile inflection point, and it is found such that the desired temperature peaking factor, $\nu_T$, is achieved. The pedestal top temperature is then found by matching the volume-average temperature, $\left<T\right>$, of the POPCON point that is being evaluated. Finally, the density profile is determined in the opposite way. Given the same profile inflection point $x_{inf}$ that was found for the temperature profile, the core inverse normalized density gradient scale length, $a/L_n$, is found (also as a single scalar value representative of the plasma core) that is consistent with the density peaking, $\nu_n$.The value of the pedestal top density is determined such that the volume-averaged density is matched. In CFSPOPCON, density peaking is set as a scalar offset on the Angioni peaking \cite{angioni_scaling_2007}, which has been shown as an appropriate approximation for SPARC \cite{rodriguez-fernandez_nonlinear_2022}. 

The choice of functional form can have a big effect on predicted performance as parabolic shapes generally place hotter and denser plasma regions in higher volume areas, as depicted in Figure \ref{subfig:profile_PRF_functionals}. Figure \ref{subfig:grad_PRF_functionals} demonstrates, however, that to sustain those parabolic profiles, large logarithmic gradients are required, which are often in excess of those expected from critical-gradient driven turbulent transport \cite{rodriguez-fernandez_predictions_2020}. As a reminder, empirical scaling laws provide no knowledge on the structure of the profiles and therefore the two example cases shown in Figure \ref{fig:PRF_functionals} have the same corresponding energy confinement time ($\tau_E$) and peaking factors ($\nu_n, \nu_T$).

\subsection{Assumption distributions} \label{sub_sec:Assumption_distributions}

To explore the effect of uncertainties we perform a Monte Carlo analysis using the confinement, profile, and impurity concentration uncertainties, which are listed in Table \ref{table:assumptions}. The means are the default values for the SPARC PRD in CFSPOPCON \cite{body_sparc_2023}, although some, like the $T_i/T_e$ ratio, differ from physics-based predictions \cite{rodriguez-fernandez_nonlinear_2022}. The standard deviations for $H$, $k_{LH}$, $\nu_{ne}$, and $\nu_{ni}$ are determined based on their respective scaling law uncertainties \cite{iter_physics_expert_group_on_confinement_and_transport_chapter_1999, martin_power_2008, angioni_scaling_2007} . The $T_i/T_e$, $\nu_T$, and $a/L_T$ uncertainties are motivated by variations in physics-based modeling \cite{rodriguez-fernandez_nonlinear_2022}. The impurity concentration uncertainties are  based on commonly observed uncertainties on high-Z, low-Z, and helium-3 minority species \cite{fontes_review_2009}. The distribution's absolute standard deviation ($\sigma$) is the relative standard deviation ($\sigma_{rel}$) multiplied by its mean ($\mu$).

\begin{table}[h]
    \centering
    \caption{Key assumption parameters that go into the POPCON modeling of the SPARC PRD. The standard deviation relative to the mean defines the distribution used for Monte Carlo sampling.}
    \begin{tabular}{cccc}
        \toprule
        Parameter & Shorthand & Mean & Relative std \\
        \midrule
        Confinement time factor & $H$ & 1.0 & 0.15\\
        Ion to electron temperature ratio & $T_i/T_e$ & 1.0 & 0.10 \\
        Electron density peaking offset & $+\nu_{ne}$ & -0.1 & 1.5 \\
        Ion density peaking offset & $+\nu_{ni}$ & -0.2 & 0.75 \\
        Temperature peaking & $\nu_{T}$ & 2.5 & 0.10 \\
        Confinement threshold factor & $k_{LH}$ & 1.0 & 0.30 \\
        Core inverse normalized gradient scale length & $a/L_T$ & 2.5 & 0.25 \\
        Helium concentration & He & 6E-2 & 0.15 \\
        Oxygen concentration & O & 3.1E-3 & 0.15 \\
        Tungsten concentration & W & 1.5E-5 & 1.0 \\
        \bottomrule
    \label{table:assumptions}
    \end{tabular}
\end{table}

\begin{figure}
    \centering
    \includegraphics[width=\linewidth]{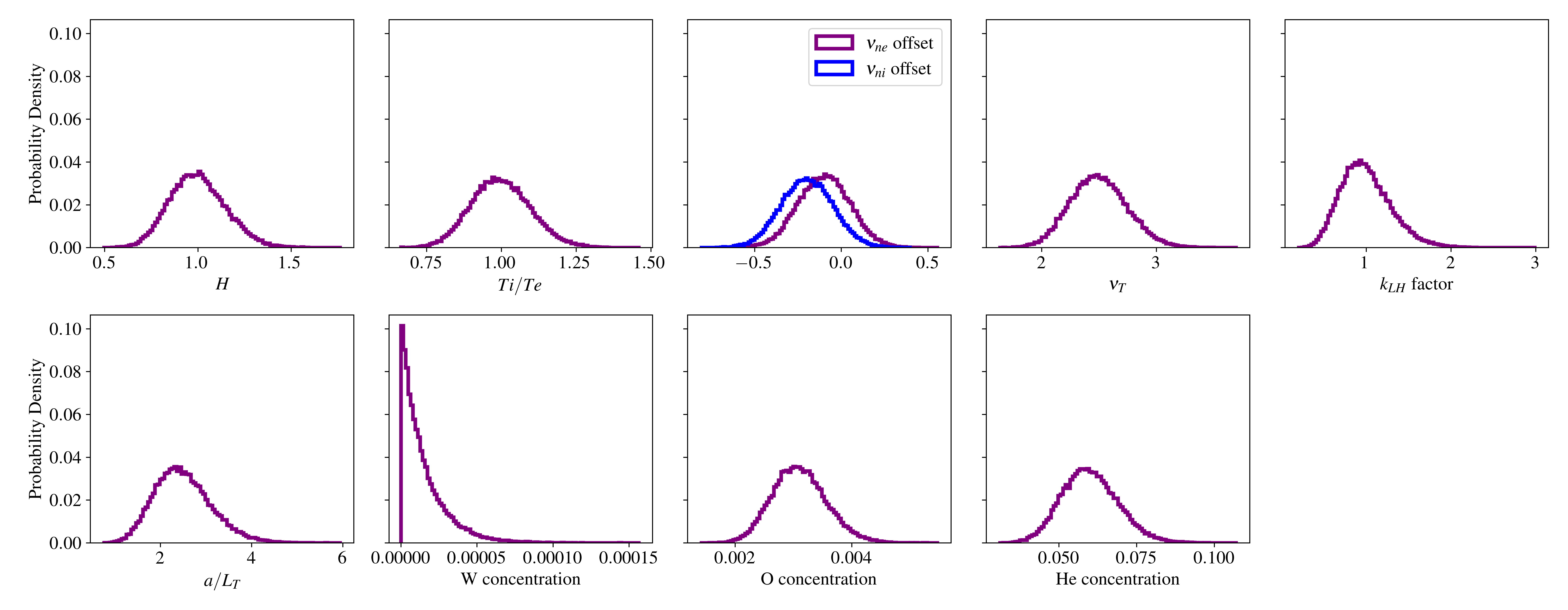}
    \caption{Distributions drawn from for Monte Carlo analysis. The offsets are Gaussian distributions and the rest are gamma distributions. Their means and standard deviations are shown in Table \ref{table:assumptions}.}
    \label{fig:assumption_dists}
\end{figure}

The Monte Carlo samples are drawn from gamma distributions, except for in the case of the offsets which are Gaussian distributions, with the mean and standard deviation relative to the mean as listed in Table \ref{table:assumptions} and shown in Figure \ref{fig:assumption_dists}. The shape parameter is calculated as $\frac{1}{\sigma_{rel}^2}$, and the scale parameter is calculated as $\mu \sigma_{rel}^2$. The samples are drawn independently from the distributions; it is assumed there is no correlation between parameters. The authors acknowledge the caveats of assuming uncorrelated parameters (particularly for transport-related quantities) but further analysis is left for future work.

\subsection{Success conditions} \label{sub_sec:Success_conditions}

In this work, we shift from a traditional POPCON approach which maximizes fusion gain to one that maximizes the probability of pointwise, in volume-averaged density - volume-averaged temperature space, compliance with success conditions given our model uncertainties. We define success conditions for the SPARC PRD as follows: 
\begin{itemize}
    \setlength\itemsep{1pt }
    \item Fusion gain $>$ 2, following the stated goal for SPARC \cite{creely_overview_2020}. Here, fusion gain is defined as $Q = P_{fus}/P_{launched}$. We assume only 90\% of RF power launched is coupled to the plasma.  
    \item Auxiliary power $<$ 25 MW, given the maximum power available on SPARC via ion cyclotron heating \cite{lin_physics_2020}.
    \item Flattop time $>$ 0s.
    \item Greenwald fraction $<$ 1, to help avoid disruptions \cite{greenwald_new_1988}.
\end{itemize}
 Note, a fusion power less than 140 MW is enforced by automatic adjustments to the relative concentrations of deuterium to tritium. Confinement mode access is enforced by choosing the H-mode or L-mode confinement scaling law as consistent with the ratio of the scrape-off-layer power to the H-mode power threshold, as discussed in the introduction to Section \ref{sec:Monte_Carlo_Methods}.

\subsection{Monte Carlo Resolution} \label{sub_sec:Monte_Carlo_Resolution}

\begin{figure}[h]
        \centering
        \includegraphics[width=\linewidth]{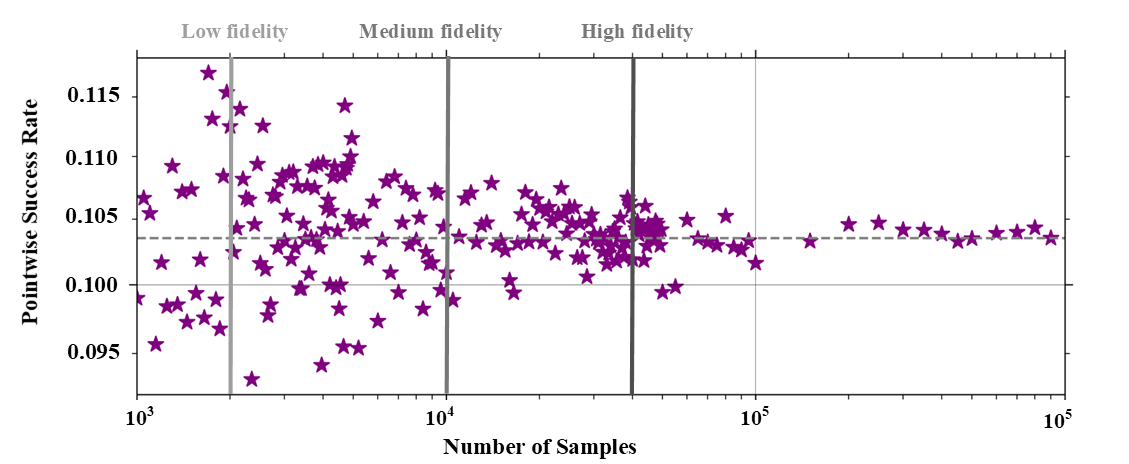}
        \caption{Scan of predicted pointwise success rate for one operating point in the statistical POPCON with increasing Monte Carlo resolution. Low, medium, and high fidelity run locations are shown. The relative errors of the fidelities are estimated from this figure.}
        \label{fig:mc_scan_success}
\end{figure}

To determine the appropriate Monte Carlo resolution a scan was performed for one operating point, as is shown in Figure \ref{fig:mc_scan_success}. From it, we define three resolutions used in this work. Low resolution is 2,000 Monte Carlo samples, which has an error of approximately 5\% when compared to the maximum resolution considered in this work. Medium resolution is 10,000 Monte Carlo samples with an error of approximately 2.5\%. High resolution is 40,000 Monte Carlo samples with an error of approximately 1\%. These values are calculated by assuming the range of values seen in Figure \ref{fig:mc_scan_success} at each resolution represents a two standard deviation variation.

\section{Statistical POPCON for SPARC Primary Reference Discharge} \label{sec:Statistical_POPCON_Results}

\begin{figure}
    \centering
    \includegraphics[width=0.9\linewidth]{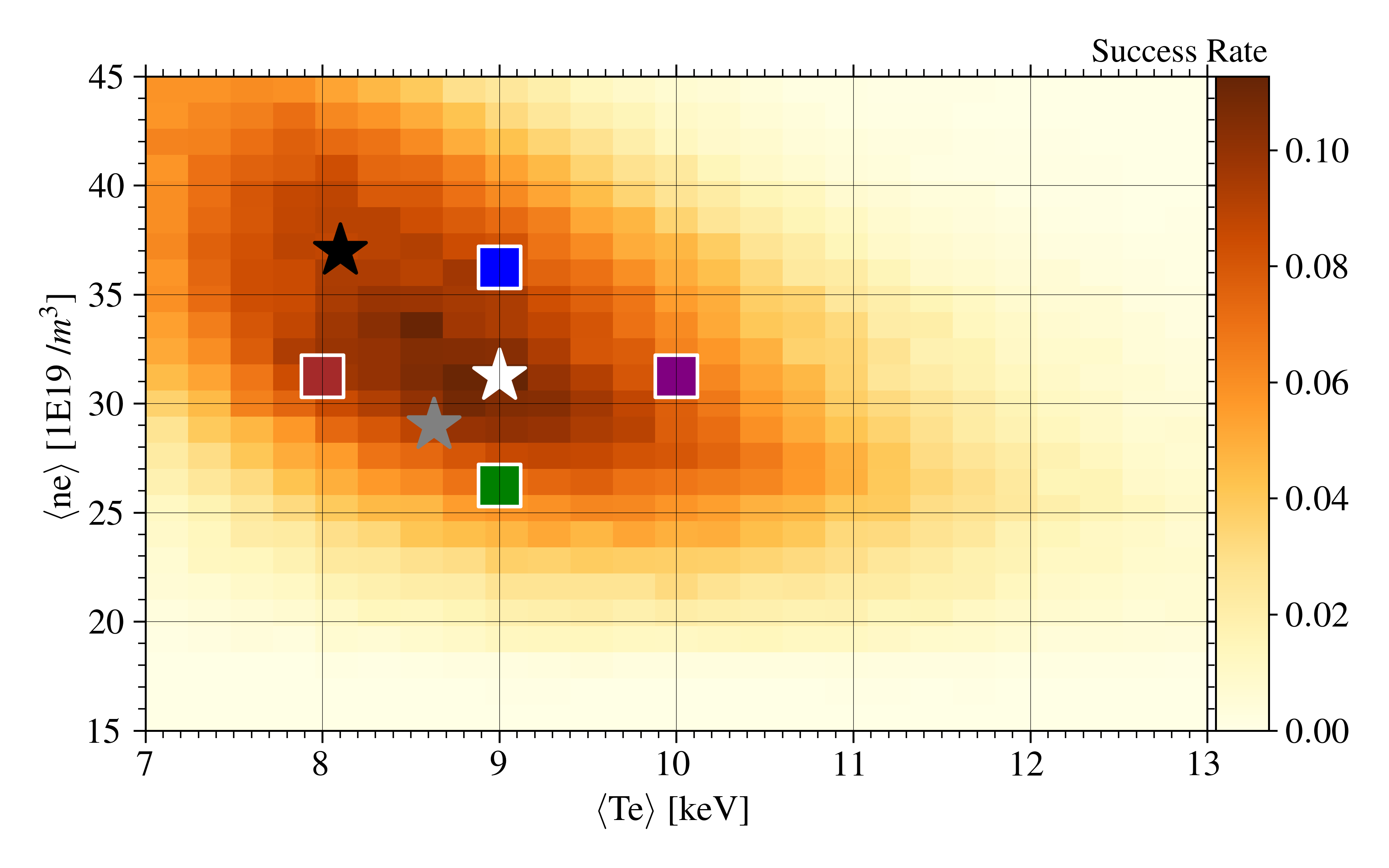}
    \caption{Statistical POPCON of the SPARC PRD with nominal assumptions and success conditions. The white star is the location of the maximum pointwise success rate. The gray star is the nominal PRD location as defined in \cite{body_sparc_2023}. The black star is where a deterministic POPCON would put the PRD when Argon used in the edge is included in the core dilution. The four squares correspond to the locations of stacked histograms shown in Figure \ref{fig:stacked_histogram_positional}. The axes are volume-averaged electron density versus volume-averaged electron temperature. The color bar represents the pointwise rate of success at each given operating point.}
    \label{fig:statistical_popcon}
\end{figure}

\begin{figure}
    \centering
    \includegraphics[width=\linewidth]{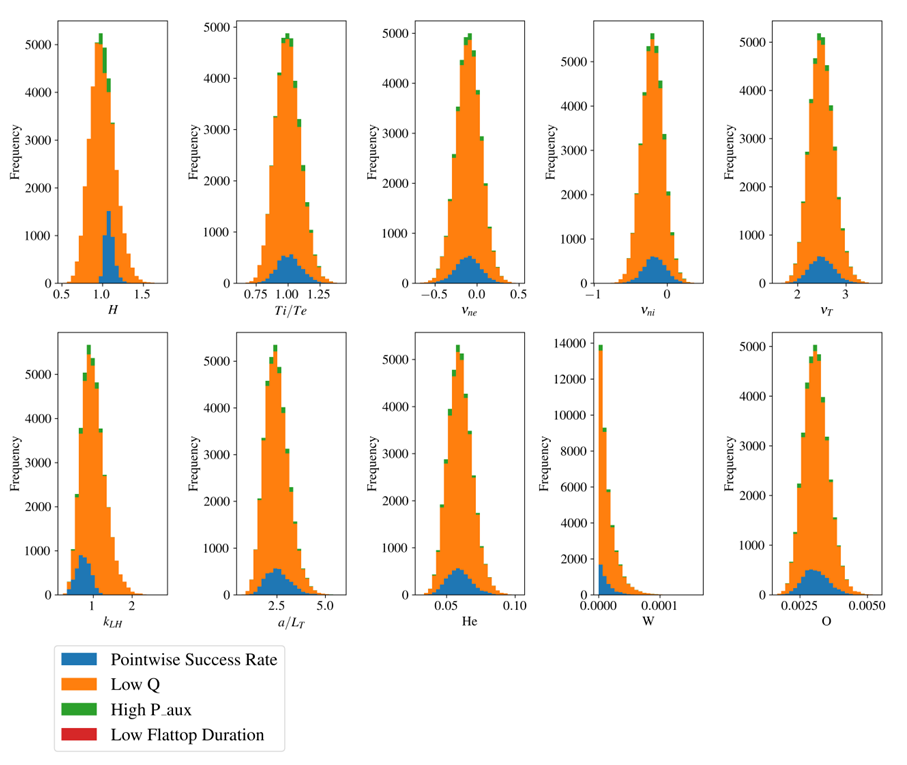}
    \caption{Stacked histogram of success condition violated versus the value of each assumption parameter on the x-axis. All parameters are varied according to their default distributions as defined in Table \ref{table:assumptions}, and these figures show the cross section of Monte Carlo sample frequency versus the value of individual parameters. Successful samples are shown in blue and cases which fail are shown with the color of the first requirement that is not met moving down the list of the legend. For example, areas that are shaded green represent samples which have a Q$>$2 but would need more than 25MW of auxiliary power to be achieved. These histograms are drawn at the location of the white star in Figure \ref{fig:statistical_popcon}. }
    \label{fig:all_stacked_histogram}
\end{figure}

\begin{figure}[htbp]    
    \centering
    \includegraphics[width=\linewidth]{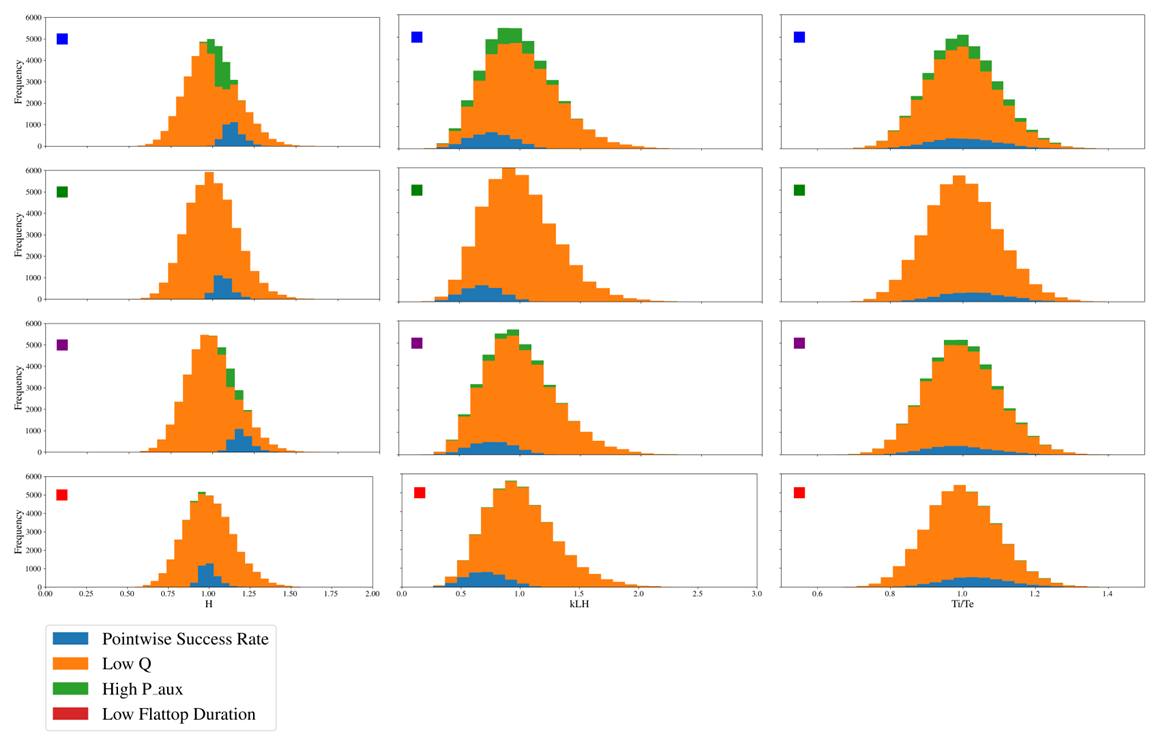}
    \caption{Variation of stacked histograms moving around density-temperature space. The color squares correspond to the locations shown in Figure \ref{fig:statistical_popcon} and the pointwise success rates as a function of the Monte Carlo sample's value of $H$, $k_{LH}$, $T_i/T_e$ are shown.  Successful samples are shown in blue and cases which fail are shown with the color of the first requirement that is not met moving down the list of the legend.}
     \label{fig:stacked_histogram_positional}
\end{figure}

A wide region of operating space was scanned in a brute force approach to produce a ``statistical POPCON."  The fraction of Monte Carlo samples which meet all the success conditions at a given operating point is referred to as the ``pointwise success rate." A broad nearly maximal zone is found for the nominal values of the assumption parameters (Table \ref{table:assumptions}) and the success conditions (Section \ref{sub_sec:Success_conditions}), as is shown in Figure \ref{fig:statistical_popcon}. The location of the maximum pointwise success rate, depicted with a white star, is relatively far from where the highest performance operating point on a deterministic POPCON that includes the impacts of core dilution from edge Argon is located, depicted with a black star. Coincidentally, the maximum pointwise success rate is relatively close to the nominal PRD location as defined in \cite{body_sparc_2023}, in which the impacts of edge Argon on the core are not considered. The assumption parameters which have the strongest impact on the pointwise success rate are $H$, $k_{LH}$, and $T_i/T_e$, as can be seen by the skewed distributions of the successful cases in Figure \ref{fig:all_stacked_histogram} and Figure \ref{fig:stacked_histogram_positional}. Additional analysis, not shown here, producing statistical POPCONs with one uncertain parameter at a time shows that 
changes in $T_i/T_e$ can also independently impact whether a Monte Carlo sample is successful.

The statistical POPCON analysis together with the stacked histograms of successful and unsuccessful cases provide important insights as to which parameters, assumptions, and criteria have the strongest effects. First, a lower L-H power threshold (represented by a smaller $k_{LH}$ multiplicative factor on the scaling law) leads to a higher rate of successful points. Monte Carlo samples that result in a scrape-off-layer power predicted to be smaller than the L-H transition threshold power use the L-mode energy confinement time scaling law making a fusion gain greater than two difficult to achieve. Importantly (and somewhat counterintuitively), only moderately high energy confinement times (represented by the $H$ multiplicative factor on the scaling law) are preferred. For too high of values, the predicted input power decreases to the point the predicted scrape-off-layer power is below the L-H threshold. Experimentally, it is seen that plasmas that are just over the L-H power threshold rarely have especially good H-mode confinement \cite{hughes_power_2011}.  In this statistical POPCON framework, we similarly reject cases with a high $H$ factor that have too low of a scrape-off-layer power (below the L-H power threshold). However, this is an example of a case where a tokamak operator could potentially choose to adjust to a higher auxiliary power, high temperature operating point and successfully achieve their goals. Temperature ratios slightly larger than one are preferred since the fusion power is directly dependent on the ion temperature. The values of $H$, $k_{LH}$, and $T_i/T_e$ that lead to the highest success rate in our model vary from one operating point to another, as shown in Figure \ref{fig:stacked_histogram_positional}, depending on the dominant failure mechanism.

For a given Monte Carlo sample, there is a higher probability that a successful operating point exists somewhere in operating space than that it is successful at a particular operating point. This captures the reality that tokamak operators have the ability to adjust the density and input power between shots. To get a ``total success rate", the number of Monte Carlo samples that are successful anywhere in operating space is divided by the total number of Monte Carlo samples tried. To make this feasible, sampling is done slightly differently. Instead of choosing a location in operating space and then drawing the desired number of Monte Carlo samples, here, we first draw the desired number of Monte Carlo samples and then evaluate same samples at all locations. A 50x50 grid is used to sample operating densities from $15\times10^{19}/m^3$ to $60\times10^{19}/m^3$ and temperatures from 5 keV to 13 keV. For the SPARC PRD statistical POPCON shown in Figure \ref{fig:statistical_popcon}, the total success rate is above $50\%$ for the set of nominal assumptions and distributions chosen in this work, as described in Section \ref{sec:Monte_Carlo_Methods}. Throughout the remainder of this work we will look at the maximum pointwise success rate instead of the total success rate because it is much easier to compute and provides information on where in operating space becomes more favorable with different changes. The maximum pointwise success rate can be thought of as a loosely proportional proxy for the total success rate. The total success rates are shown in Table \ref{table:success_rates} for limits of the scans on key distribution means, seen in Figure \ref{fig:means_scan}, and on the edge argon enrichment, seen in Figure \ref{fig:Ar_enrichment}. The details of these scans will be discussed further in section \ref{sec:Scanning_Results}. Here, we note that for small changes in pointwise success rate, there is a similar change in the relative value of the total success rate.

\begin{table}[h]
    \centering
    \caption{Total success rates with adjustments to the mean value of key distributions as in Figure \ref{fig:means_scan} and edge argon enrichment as in Figure \ref{fig:Ar_enrichment}.}
    \begin{tabular}{lcc}
        \toprule
        Adjustment & \multicolumn{2}{c}{Total Success Rate} \\
        \midrule
        Baseline & \multicolumn{2}{c}{0.58} \\
        \midrule
        & Min & Max \\
        $H$ mean: 0.85 - 1.15 & 0.33 & 0.72 \\
        $k_{LH}$ mean: 0.7-1.3 & 0.85 & 0.29 \\
        $T_i/T_e$ mean: 0.90-1.10 & 0.50 & 0.63 \\
        Edge Ar enrichment: 2-10 & 0.23 & 0.76\\
        \bottomrule
    \end{tabular}
    \label{table:success_rates}
\end{table}

\begin{figure}[h]
    \centering
    \includegraphics[width=0.49\linewidth]{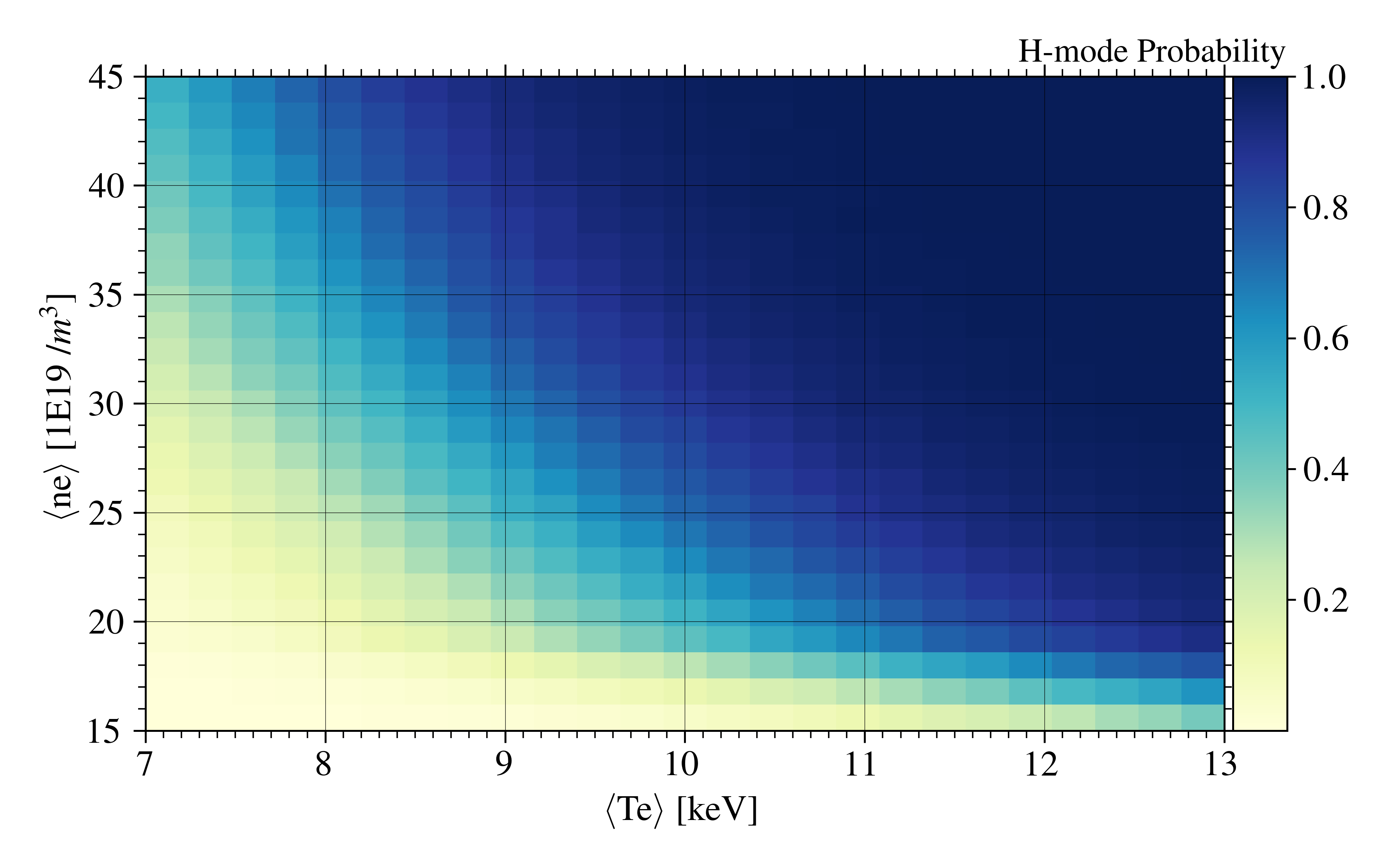}
    \caption{Probability of H-mode confinement time scaling law \cite{yushmanov_scalings_1990} being used at each operating point in accordance with the fraction of points that are about the L-H transition power threshold \cite{martin_power_2008}. volume-averaged density vs volume-averaged temperature coordinates are used.}
    \label{fig:H-mode_prob}
\end{figure}

Moving to temperatures lower than the optimum, as shown in the histograms with the red square in Figure \ref{fig:stacked_histogram_positional},  causes more Monte Carlo samples to be below the L-H transition threshold power and thus to be evaluated using the less favorable L-mode energy confinement time scaling law. The reduced likelihood of H-mode access is shown in Figure \ref{fig:H-mode_prob}.  At this lower temperature, H factors of one or less are preferred, as seen in Figure \ref{fig:stacked_histogram_positional}, due to the resulting increased scrape-off-layer power. At too low of densities, as shown by the green square in Figure \ref{fig:stacked_histogram_positional}, the probability of having access to H-mode confinement also decreases, as shown in Figure \ref{fig:H-mode_prob}. Moving to higher temperatures, as shown by the histograms with the purple square in Figure \ref{fig:stacked_histogram_positional}, causes the fusion gain to decrease because more input power is required. This is readily apparent on the deterministic POPCONs presented in \cite{body_sparc_2023}. The decreasing energy confinement time with power is a result of the power degradation factor in all of the energy confinement scaling laws. Similarly, at too high of densities, as shown in the histograms with the blue square in Figure \ref{fig:stacked_histogram_positional}, more input power than available would be needed to maintain the high stored energy of the plasma. 

\begin{figure}
    \centering
    \includegraphics[width=0.5\linewidth]{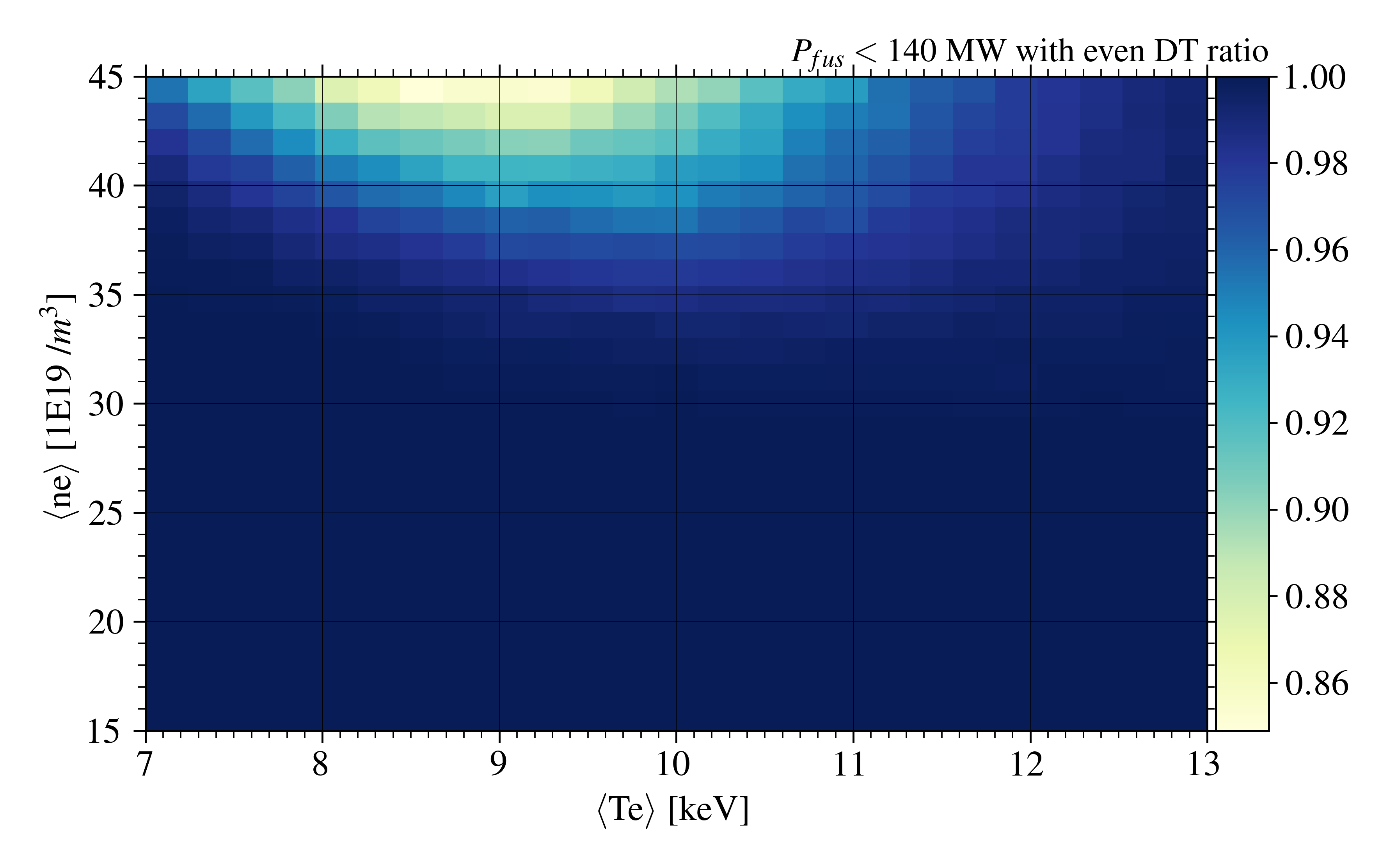}
    \caption{Probability of a fusion power less than 140 MW being achieved with an equal concentration of deuterium and tritium. Note, the high density, high temperature upper right corner does not require the D-T ratio to be adjusted because the fusion power is already limited by the amount of argon required to be injected to limit the scrape-off-layer power to a reasonable level. Volume averaged density vs volume averaged temperature coordinates are used.}
    \label{fig:DT_ratio}
\end{figure}

The automated enforcement of the 140 MW fusion power limit can be sen in Figure \ref{fig:DT_ratio}. Up to 10\% of the highest performing samples per operating point exceed the fusion power of 140 MW and are brought back to the power limit by increasing the relative concentration of deuterium to tritium. If, with this adjusted D-T ratio, the Monte Carlo sample still meets the requirement of a fusion gain greater than two, it is considered successful.

\begin{figure}[h]
    \centering
    \begin{subfigure}{0.49\linewidth}
        \centering
        \includegraphics[width=\linewidth]{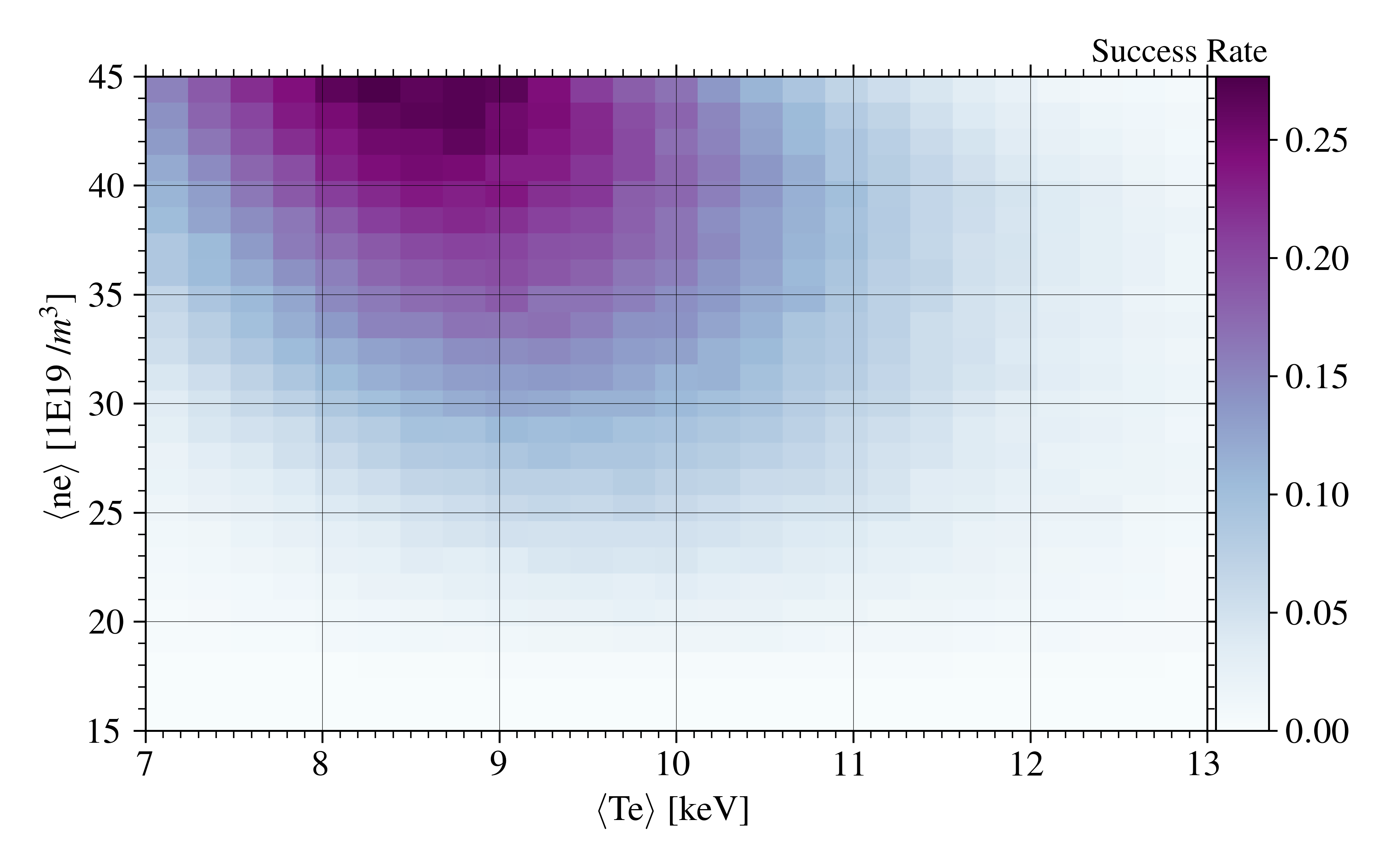}
        \caption{}
        \label{fig:Q_2_success_condition}
    \end{subfigure}
    \hfill
    \begin{subfigure}{0.49\linewidth}
        \centering
        \includegraphics[width=\linewidth]{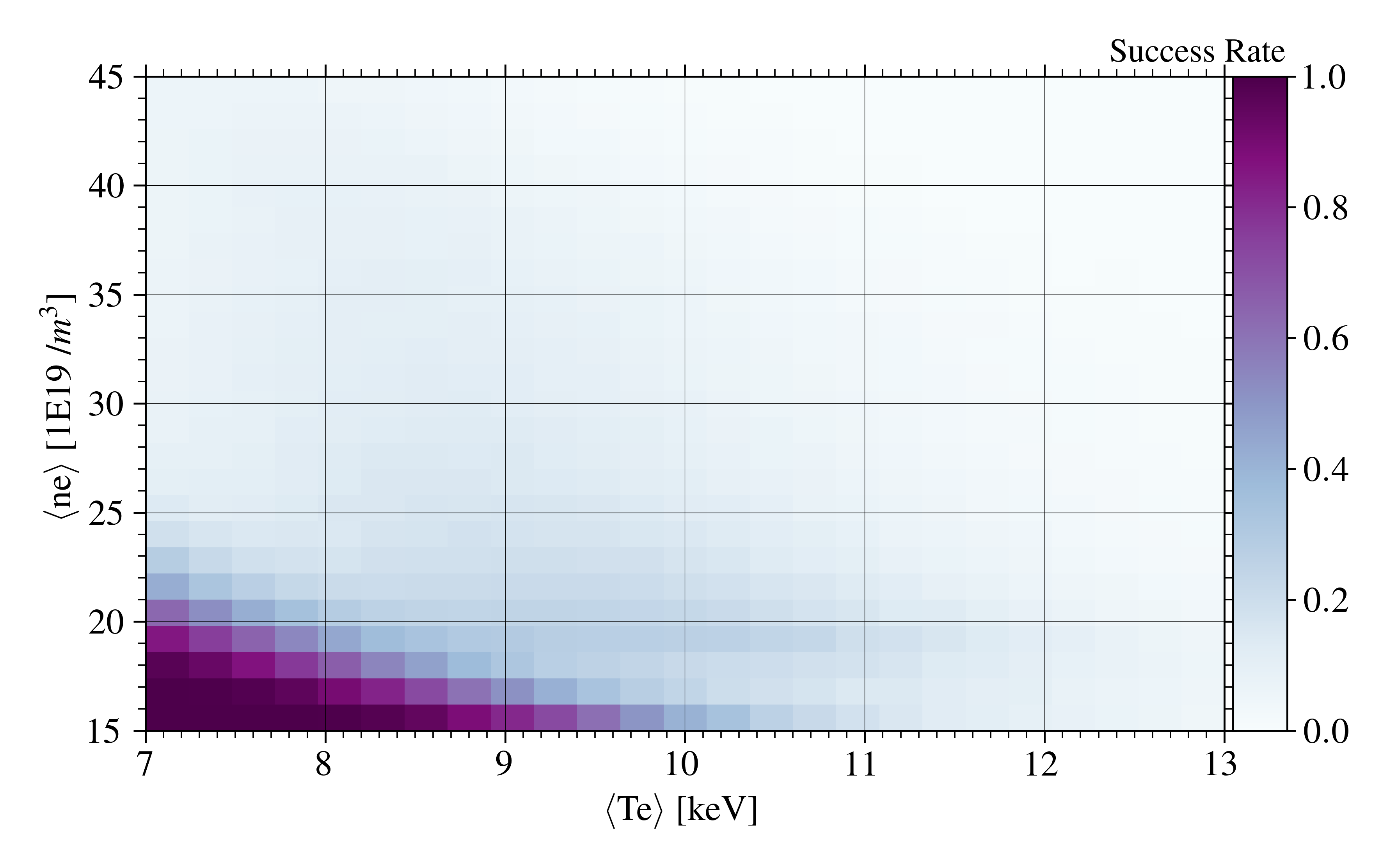}
        \caption{}
        \label{fig:P_aux_25_success_condition}
    \end{subfigure}
    \hfill
    \begin{subfigure}{0.49\linewidth}
        \centering
        \includegraphics[width=\linewidth]{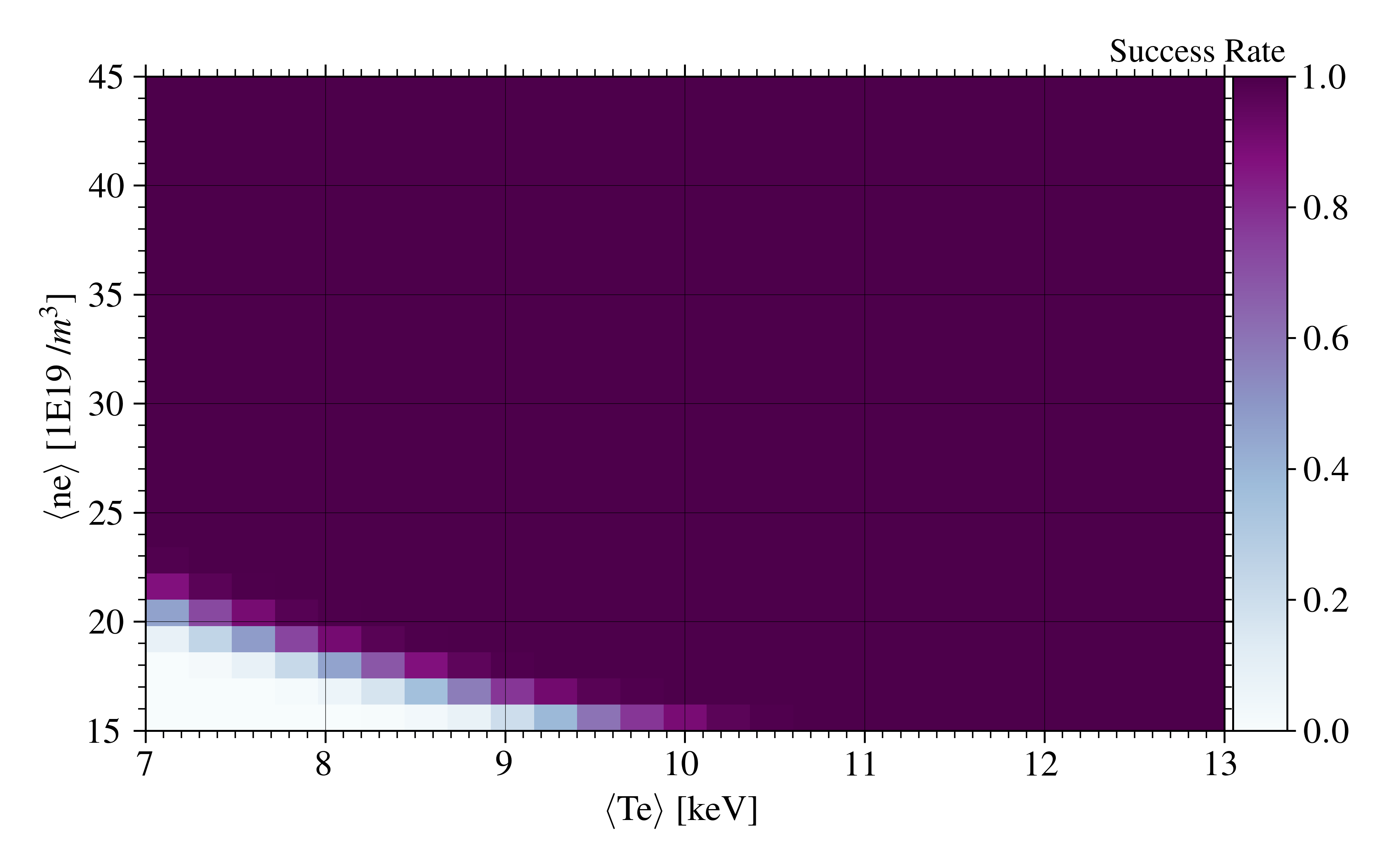}
        \caption{}
        \label{fig:t_flattop_0_success_condition}
    \end{subfigure}
        \caption{Break down of the pointwise success rates by each individual success condition. (a) Pointwise rate of achieving Q$>$2, (b) pointwise rate of achieving auxiliary power $<$ 25MW, and (c) pointwise rate of achieving a non-zero flattop period at each operating point. volume-averaged density vs volume-averaged temperature coordinates are again used.}
        \label{fig:individual_success_conditions}
\end{figure}

 Figure \ref{fig:individual_success_conditions}, we break down at which operating points each individual success criterion is most likely to be satisfied. The requirement of a fusion gain greater than two pushes the solution to higher densities and temperatures around 9 keV.  Achieving an auxiliary power less than 25 MW favors lower temperature and density solutions, which have lower stored energy. The tension between these criteria results in a mid-range solution for both density and temperature. The requirement that a flattop time exists is met in all samples except for those at extremely low density, because there are limited charge carriers, and low temperature, because resistivity increases at low temperatures. The Greenwald fraction is only 0.37 for the SPARC PRD \cite{creely_overview_2020}. Since it is still deterministic, it does not change from sample to sample, and this criterion is met for all points on plots shown in this paper.

\begin{figure}
    \centering
    \includegraphics[width=0.5\linewidth]{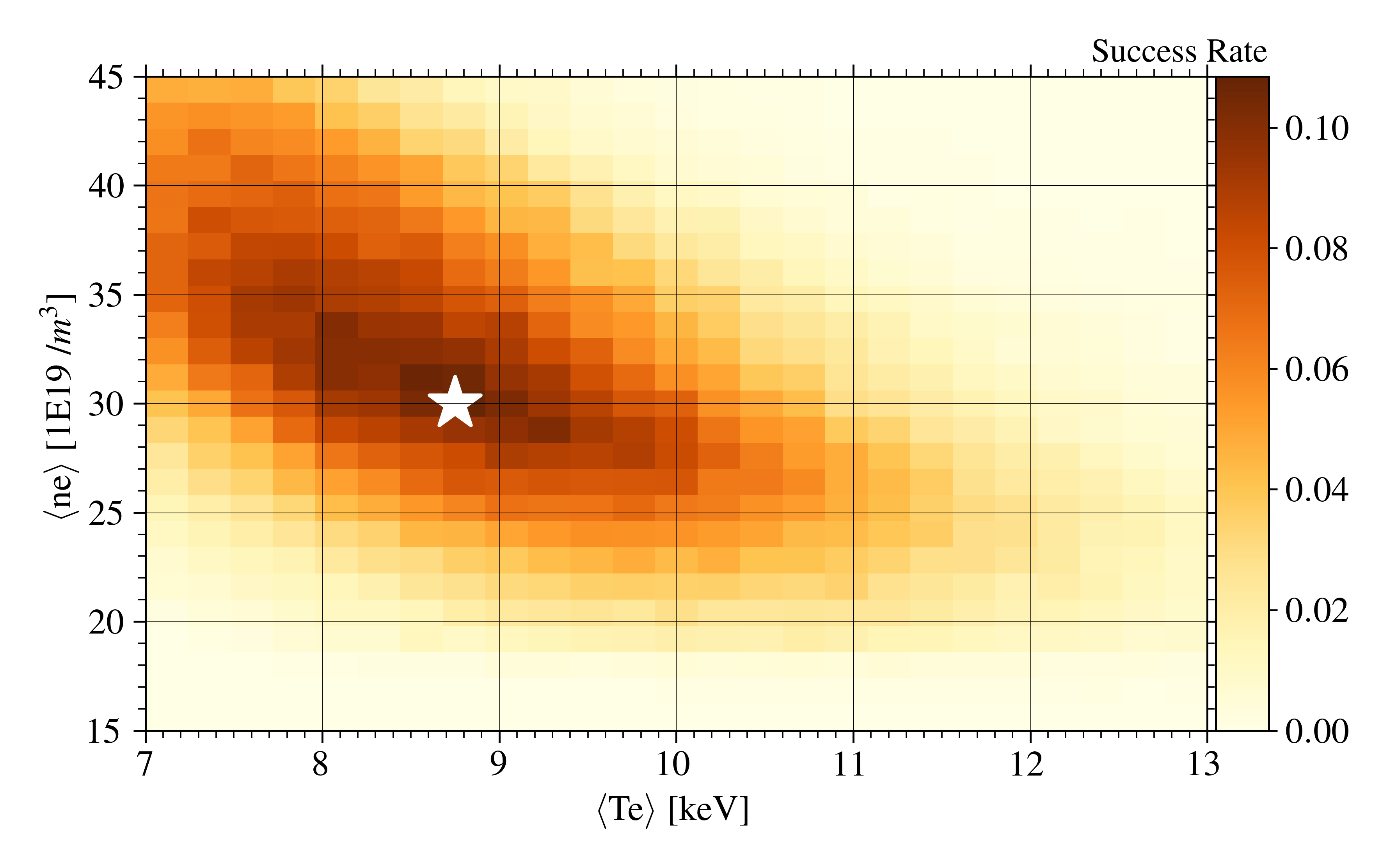}
    \caption{Statistical POPCON generated using all base assumption distributions but with the an updated H-mode confinement time scaling law from \cite{verdoolaege_updated_2021}. The white star represents the location of the maximum pointwise success rate. The axes are volume-averaged density versus volume-averaged temperature.}
    \label{fig:ITPA20_statistical_popcon}
\end{figure}

We next explore the impact of using the updated ITPA global H-mode confinement scaling laws, specifically the engineering parameter scaling law found in Table 16 of \cite{verdoolaege_updated_2021}, which includes more data from metal-walled, higher power, and higher density machines. It finds a much weaker dependence on line-averaged density. The resulting statistical POPCON can be seen in Figure \ref{fig:ITPA20_statistical_popcon}. We find the statistical POPCON generated with this new model has a very similar shape to the statistical POPCON generated with the ITER98y2 scaling \cite{yushmanov_scalings_1990},

Unfortunately, producing the statistical POPCONS discussed in this section with an operating grid of 25x25 at a Monte Carlo resolution of 10,000 is relatively expensive, on order of 10 CPU hours per statistical POPCON, which prohibits being able to further explore the impact of the assumptions made in the modeling. For this, we turn to optimization techniques, which allow us to find operating point with the maximum success rate much more quickly than a brute force search.


\section{Bayesian Optimization Methods} \label{sec:Bayesian_Optimization_Methods}

We explore two black box optimization methods beyond the brute-force grid search applied in Section \ref{sec:Statistical_POPCON_Results}: the Bayesian and Powell optimization algorithms. Powell optimization is a conjugate direction method, which is an iterative algorithm between a steepest descent algorithm and a Newton algorithm. Bayesian optimization fits surrogate functions to the black box evaluations made thus far and then uses an acquisition function to select the most valuable point to evaluate with the full black box next. For Powell optimization, we use the standard SciPy optimization library routine \cite{noauthor_minimizemethodpowell_nodate}. For Bayesian optimization we leverage the capabilities of the BoTorch framework \cite{Balandat2019} through the MITIM framework \cite{rodriguez-fernandez_pabloprfmitim-fusion_2024}, as used in PORTALS \cite{rodriguez-fernandez_enhancing_2023}. We use the Monte Carlo logarithmic expected improvement \cite{NEURIPS2023_419f72cb}, with 1024 Monte Carlo samples in the acquisition function. Our surrogates are Gaussian processes with a Matern kernel and a linear mean. Initialization is done with five training points selected according to Latin hypercube sampling. Runs are done for a fixed number of iterations with the statistical POPCON Monte Carlo resolution specified for each.

\begin{figure}[htpb]
    \centering
    \begin{subfigure}{0.75\linewidth}
        \centering
        \includegraphics[width=\linewidth]{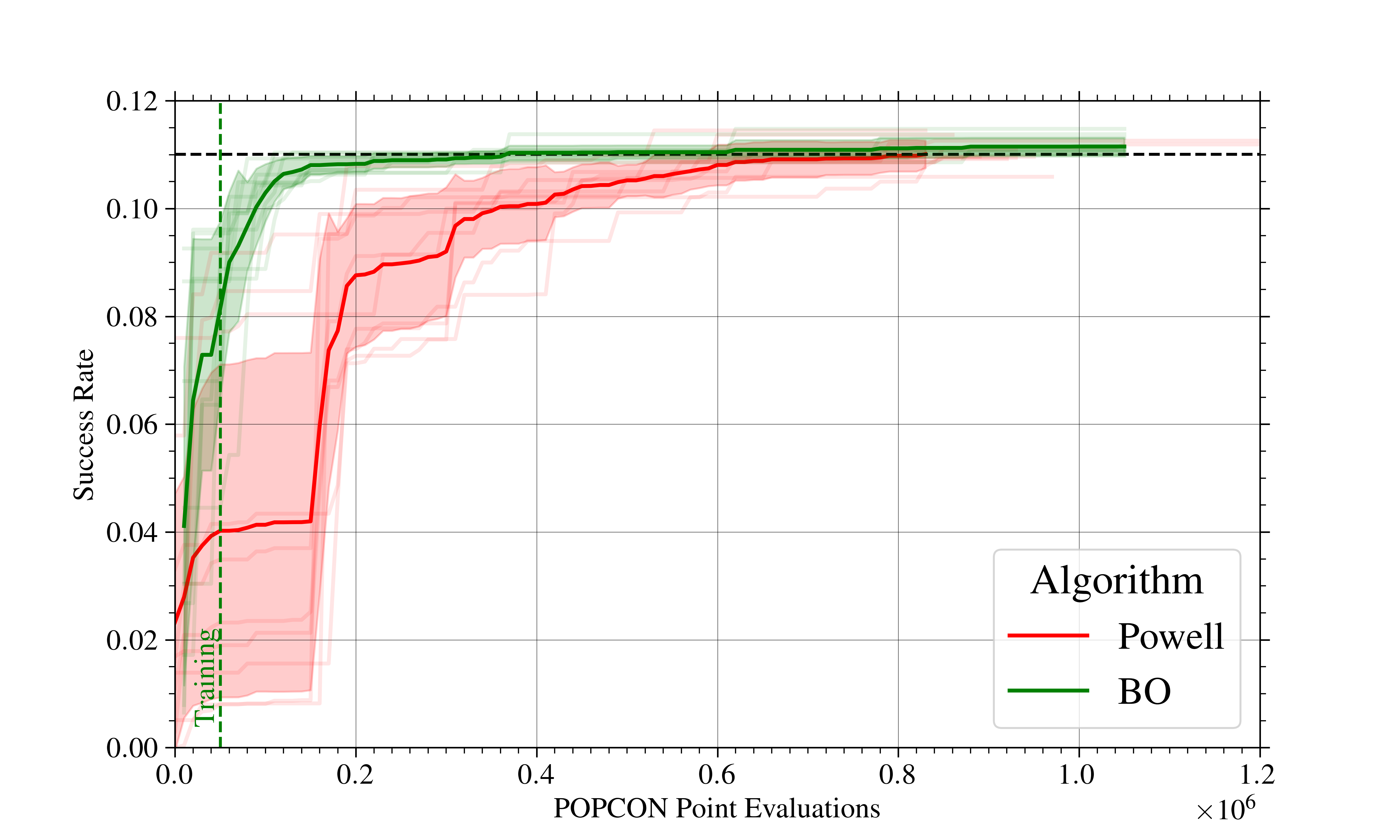}
        \caption{}
        \label{fig:Powell_vs_BO_success_rate}
    \end{subfigure}
    \hfill
    \begin{subfigure}{0.47\linewidth}
        \centering
        \includegraphics[width=\linewidth]{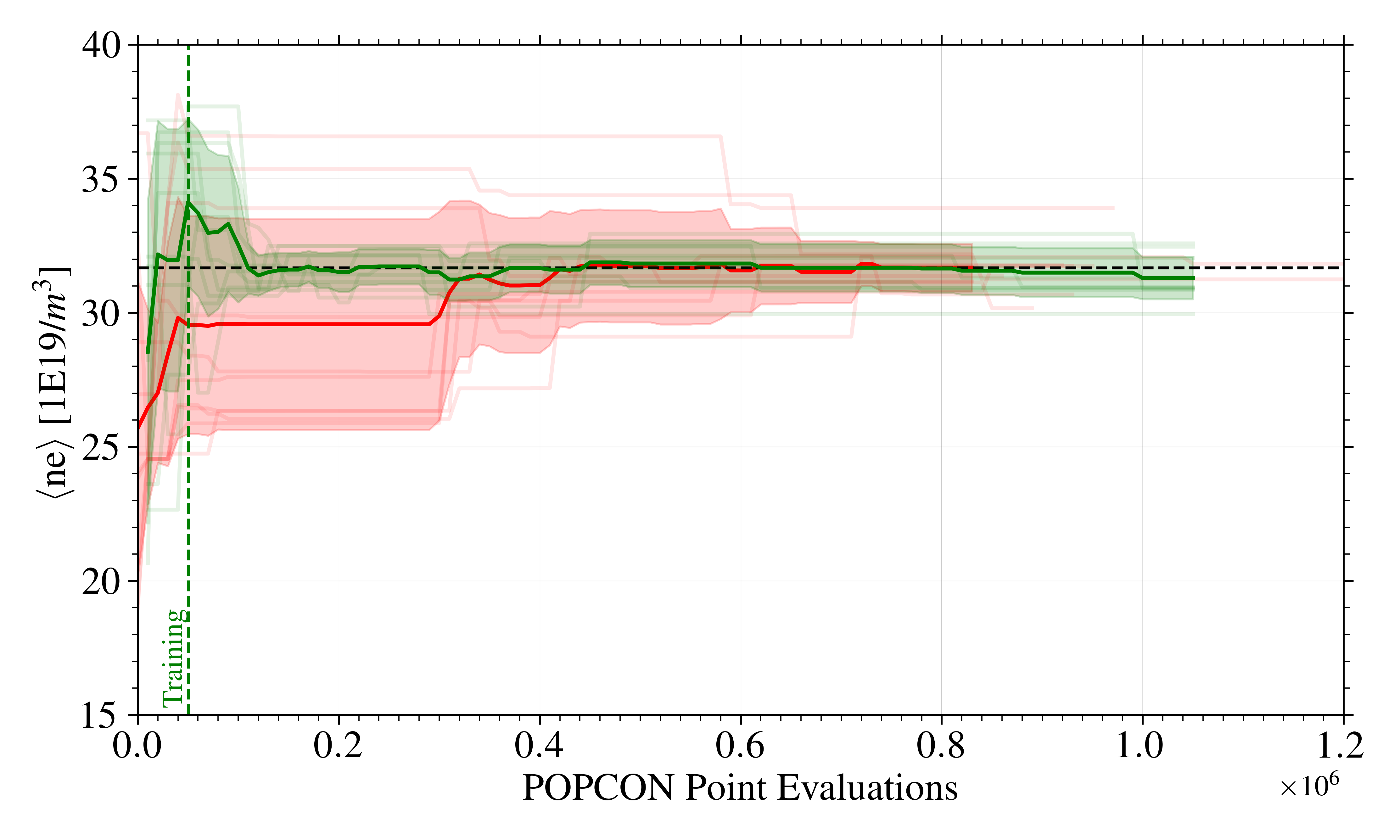}
        \caption{}
        \label{fig:Powell_vs_BO_ne}
    \end{subfigure}
    \hfill
    \begin{subfigure}{0.5\linewidth}
        \centering
        \includegraphics[width=\linewidth]{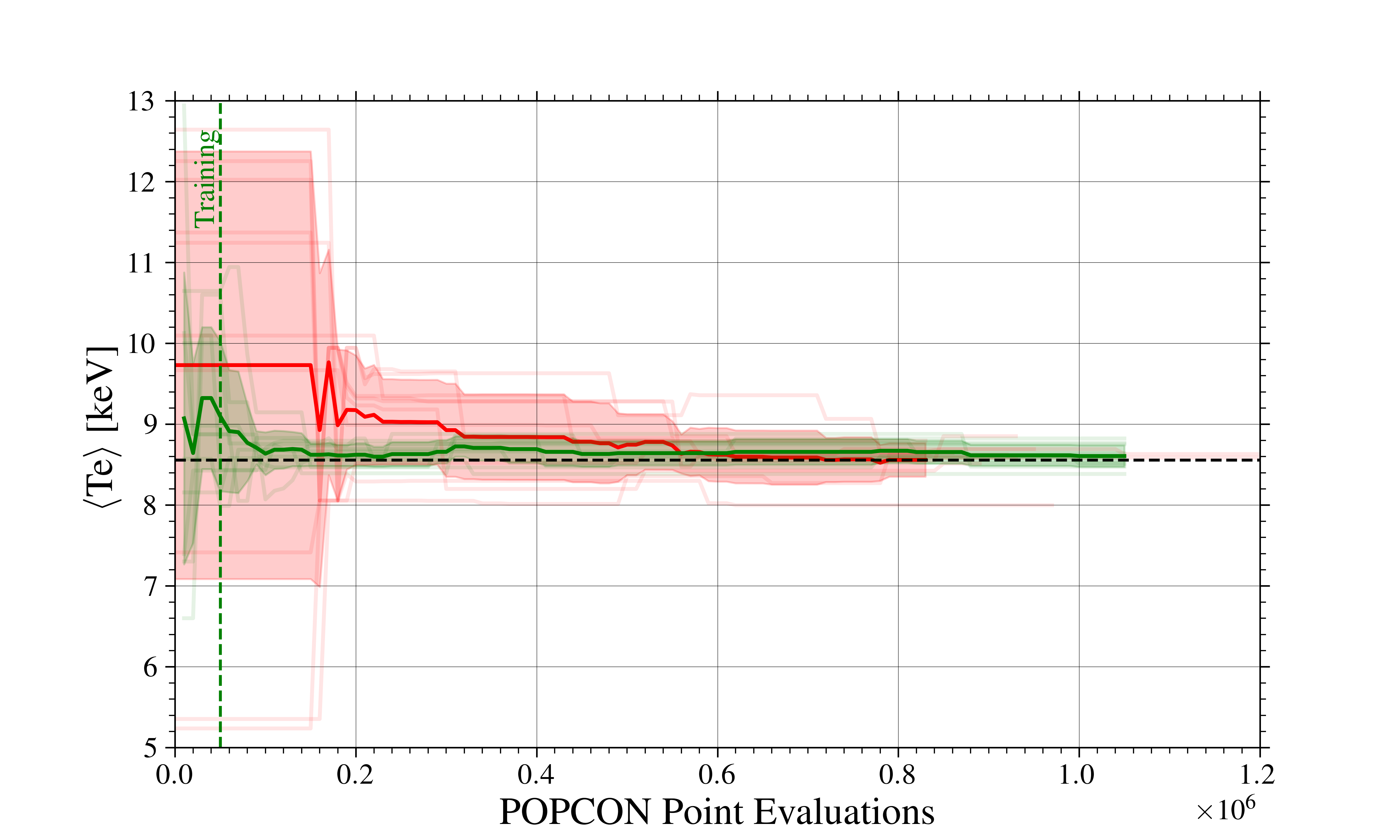}
        \caption{}
        \label{fig:Powell_vs_BO_Te}
    \end{subfigure}
    \caption{Comparison of Powell and Bayesian optimization convergence, both using medium Monte Carlo fidelity (10,000 samples). The x-axis is the number of POPCON point evaluation. (a) Shows the convergence of the maximum pointwise success rate, (b) the convergence of the volume-averaged density with maximum pointwise success rate, and (c) the convergence of the volume-averaged temperature with the maximum pointwise success rate. The vertical green dotted lines show where the where the Bayesian optimization routine switches from initial training points to optimization. The horizontal black dotted lines show the best value found in all of the Powell iterations.}
    \label{fig:Powell_vs_BO}
\end{figure}

First, we compare the efficiency of Powell optimization with Bayesian optimization for our black box  (i.e. the pointwise success rate) evaluated with medium fidelity (10,000 Monte Carlo samples). We compare their rates of convergence for 10 different random seeds. For Powell optimization, the seed determines the initial point from which the descent begins. In Bayesian optimization, the random seed controls the five initial locations the surrogates are trained on, and the starting points for the Gaussian process training and acquisition function optimization. Averaging across seeds lets us see the expected performance of each algorithm. Through its initialization with a coarse sampling of the entire operating space, Bayesian optimization, on average, successfully produces a more accurate estimation of the pointwise success rate, and location in operating space of it, for any number of evaluations, as demonstrated in Figure \ref{fig:Powell_vs_BO}.

From there, we explore the benefits of using multi-fidelity Bayesian optimization approaches. Here, we consider a multi-fidelity approach to be one that changes the number of Monte Carlo samples used per point evaluation directly as a function of the iteration number. Once a certain number of iterations are complete, the Monte Carlo sample size is set to increase. Smaller samples sizes are faster to evaluate but result in more statistical uncertainty on the value of the pointwise success rate calculated. Therefore, low-fidelity iterations use a relatively small Monte Carlo sample size. High-fidelity, later iterations use a relatively large Monte Carlo sample size.

\begin{figure}[htpb]
    \centering
    \begin{subfigure}{0.9\linewidth}
        \centering
        \includegraphics[width=\linewidth]{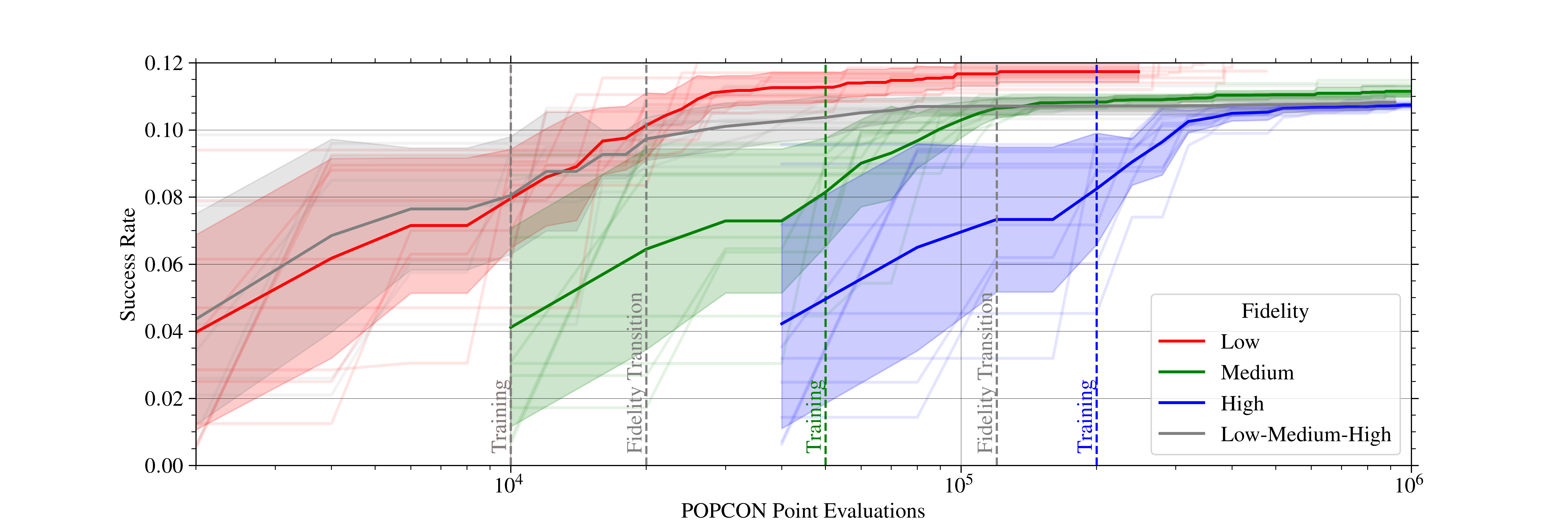}
        \caption{}
        \label{subfig:mf_success_rate}
    \end{subfigure}
    \hfill
    \begin{subfigure}{0.49\linewidth}
        \centering
        \includegraphics[width=\linewidth]{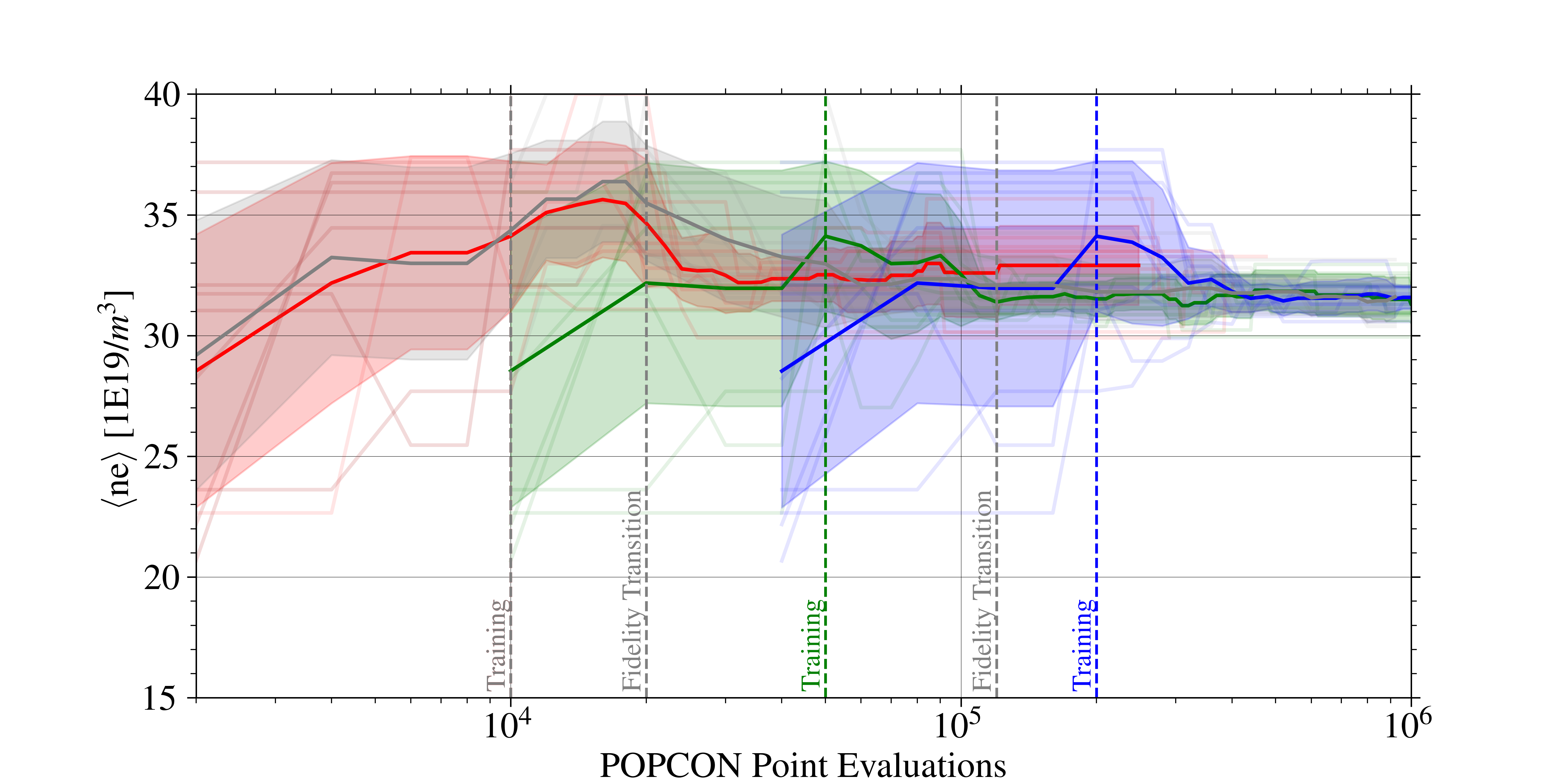}
        \caption{}
        \label{subfig:mf_ne}
    \end{subfigure}
    \hfill
    \begin{subfigure}{0.49\linewidth}
        \centering
        \includegraphics[width=\linewidth]{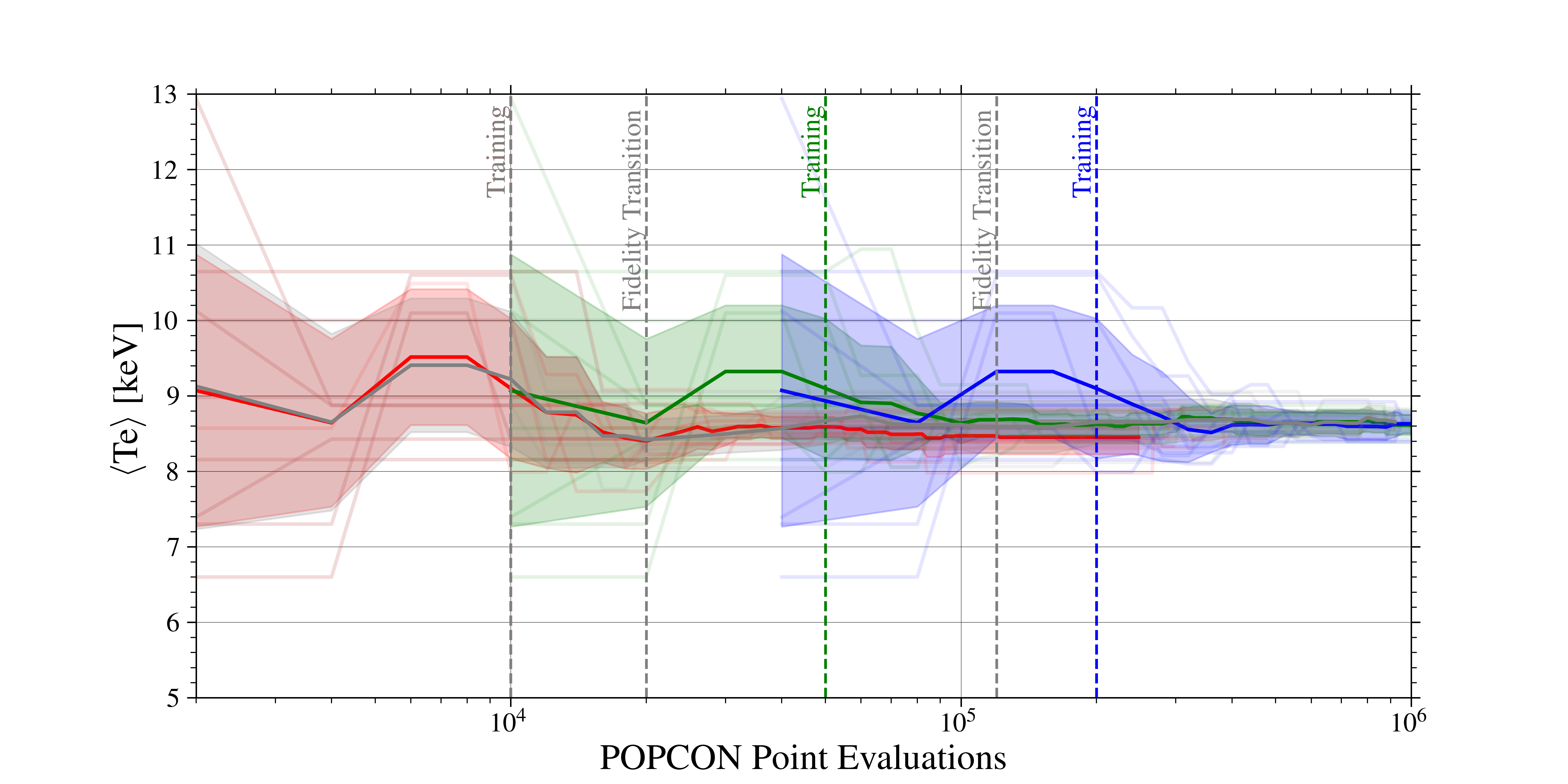}
        \caption{}
        \label{subfig:mf_Te}
    \end{subfigure}
        \caption{Comparison of the accuracy and efficiency of convergence with low (red: 2,000 samples), medium (green: 10,000 samples), and high (blue: 40,000) fidelity compared with a multi-fidelity approach (gray) that transitions between all three.  The x-axis is the number of POPCON point evaluations which increases by the number of Monte Carlo samples used in each iteration, depending on its fidelity. The convergence for each individual seed is shown in transparent lines. The average convergence of all seeds is shown with the dark lines, and the one standard deviation spread of the seeds from the mean in shown with the filled error bars. (a) Shows the convergence of the pointwise success rate, (b) the convergence of the volume-averaged density with the highest pointwise success rate, and (c) the convergence of the volume-averaged temperature with the highest pointwise success rate. Vertical dotted lines show transitions from using randomly selected training data to points selected by the optimization algorithm as well as where the transitions from low to medium and medium to high fidelity occur in the multi-fidelity approach. Both the low-fidelity and multi-fidelity cases switch from training to optimization at 10,000 point evaluations.}
        \label{fig:mf}
\end{figure}

We find lower fidelity approaches converge more quickly in number of POPCON evaluations, but the maximum pointwise success rate reported is higher than the true maximum. This is because there are statistical fluctuations in each pointwise success rate determination, and the maximum value found is shown in Figure \ref{subfig:mf_success_rate}. Additionally, the process of training surrogates and finding the optimal next point to sample takes ~30\% of the time for a low fidelity evaluation but only ~5\% of the time for a high fidelity evaluation. Therefore, higher fidelity approaches reduce the computational overhead from the optimization algorithms compared to more evaluations at low fidelity. We choose a gradual method that increases fidelity from low to medium and then to high with increasing number of iterations. The five initial training points are evaluated at low-fidelity. After five more iterations of the Bayesian optimization workflow, it switches to medium fidelity for ten more iterations before using high fidelity for the remainder evaluations. As demonstrated in Figure \ref{fig:mf}, this switching approach (shown in gray) achieves convergence as fast as the low-fidelity approach (red) but retains the accuracy of the high fidelity evaluation (blue). This is the approach we will use for all following physics-based scans. When reporting maximum pointwise success rate and optimal operating points, we will only consider the results of high-fidelity evaluations.

\section{Exploration of Model Assumptions and Engineering Parameters} \label{sec:Scanning_Results}
Leveraging multi-fidelity Bayesian optimization, we explore the effects of changing the relative standard deviations and means of the factors shown in Table \ref{table:assumptions}. Then, we explore the effect of the ratio at which impurities in the edge, leveraged for a self-consistent divertor solution, are retained in the edge instead of leaking into and diluting the core. Finally, we consider the effect potential engineering uncertainties could have on the device's performance. 

\subsection{Effect of Reducing Uncertainties} \label{subsec:Reducing_Uncertainties}

\begin{figure}[htbp]
    \centering
    \begin{subfigure}[b]{0.75\textwidth}
        \includegraphics[width=\textwidth]{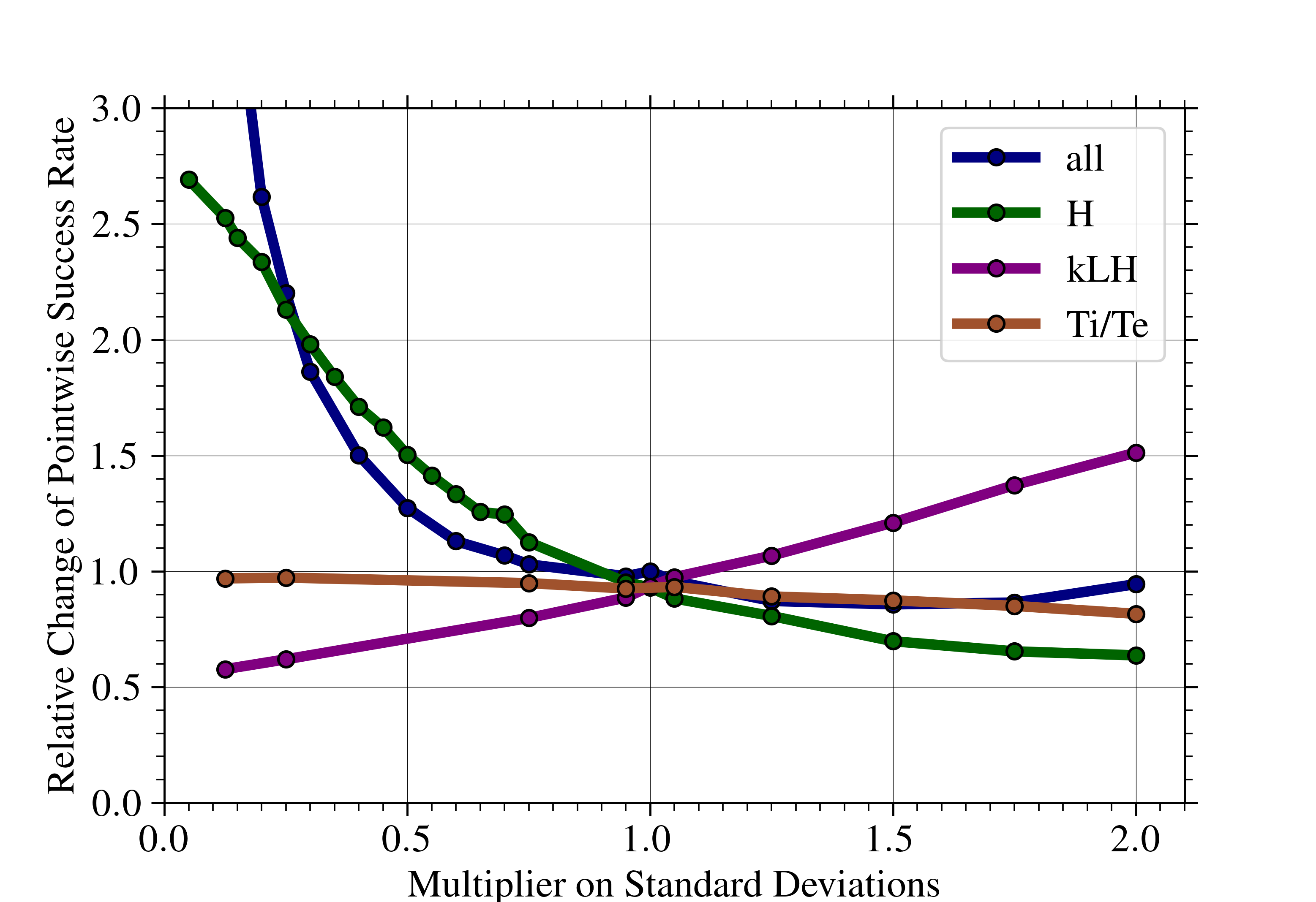}
        \caption{}
        \label{subfig:std_scan_success_rates}
    \end{subfigure} 
    \hfill
    \begin{subfigure}[b]{0.32\textwidth}
        \includegraphics[width=\textwidth]{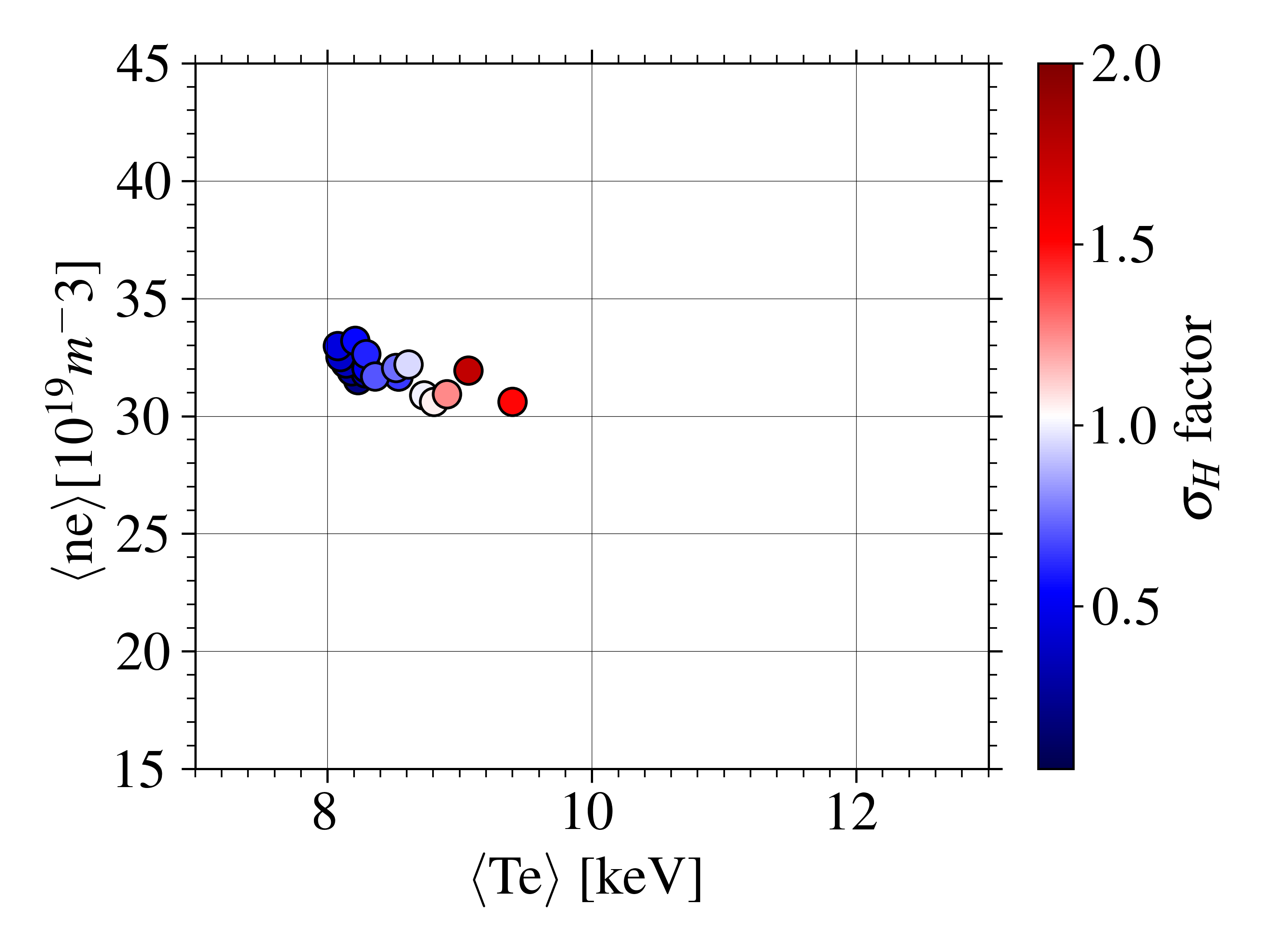}
        \caption{}
        \label{subfig:H_std_ne_Te}
    \end{subfigure} 
    \begin{subfigure}[b]{0.32\textwidth}
        \includegraphics[width=\textwidth]{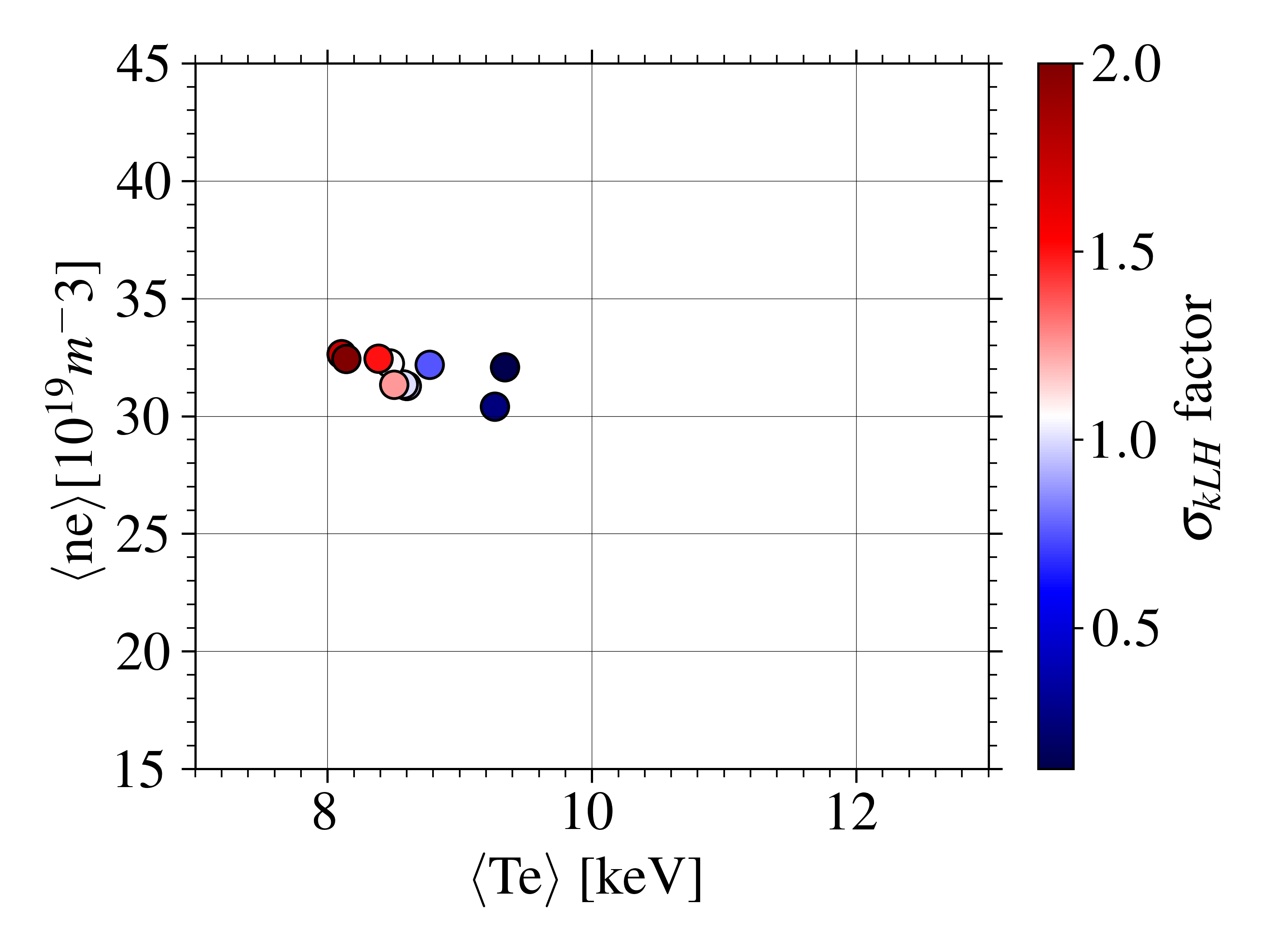}
        \caption{}
        \label{subfig:kLH_std_ne_Te}
    \end{subfigure} 
    \begin{subfigure}[b]{0.32\textwidth}
        \includegraphics[width=\textwidth]{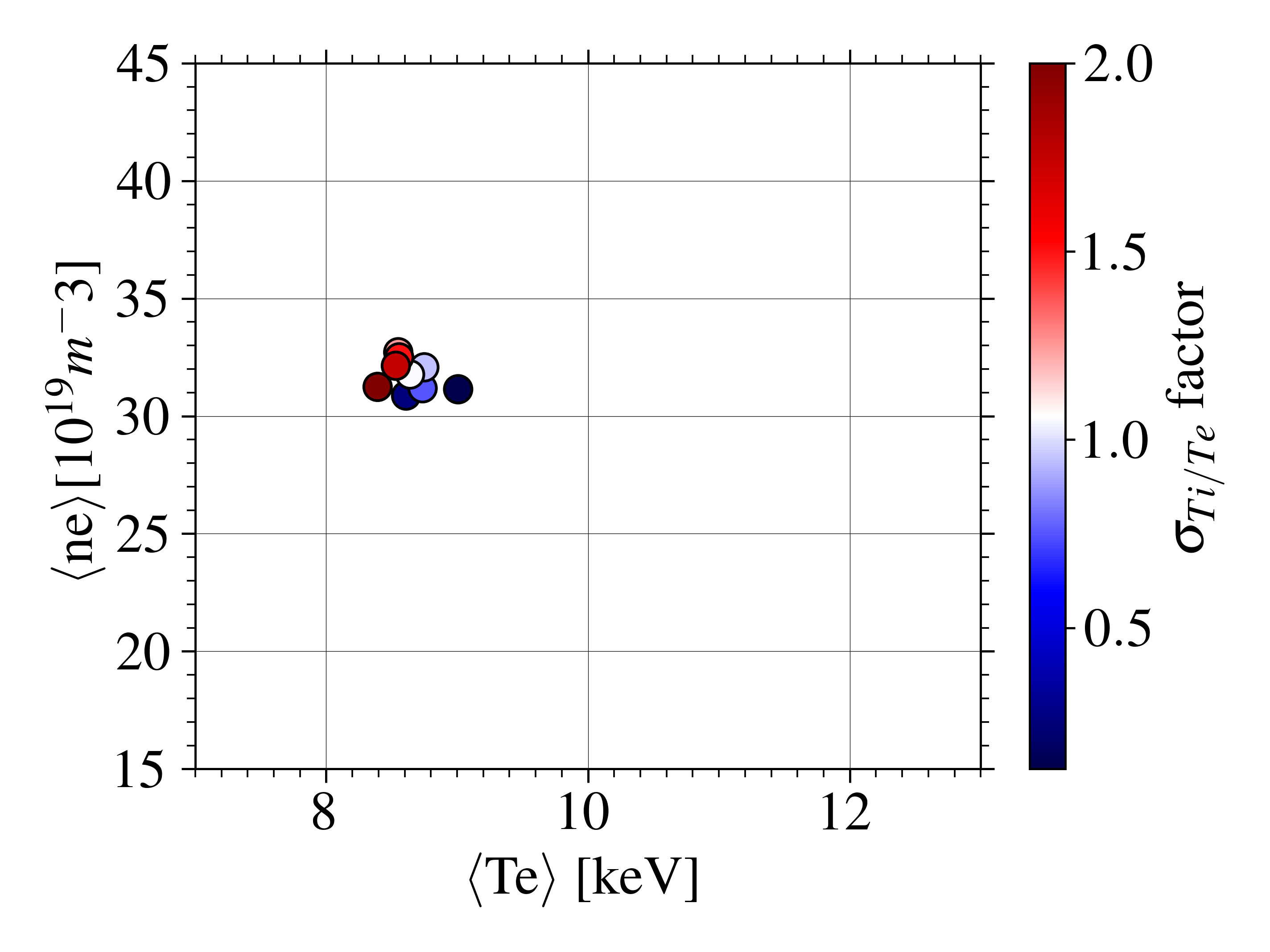}
        \caption{}
        \label{subfig:temp_ratio_std_ne_Te}
    \end{subfigure}
    \caption{Impact of increasing and decreasing relative standard deviations of assumption parameters by a constant factor while holding distribution means fixed. Figure (a) looks at the change in the maximum pointwise success rate. Figures (b-d) show the location, in volume-averaged density - volume-averaged temperature space, of the maximum pointwise success rate. The case of all uncertainties changing is explored (blue) in addition to changing the individual uncertainties of $H$ (green), $k_{LH}$ (purple), and $T_i/T_e$ (brown). Reducing our uncertainty on $H$ greatly improves SPARC's predicted performance and causes lower temperatures solutions to be favored. However, increasing certainty about $k_{LH}$ actually decreases the maximum pointwise success rate and causes higher temperature solutions to be a favored.}
    \label{fig:std_scan}
\end{figure}

First, we test how changing all the distributions' relative standard deviations by a fixed factor modifies the predicted maximum pointwise success rate. We find, as can be seen in Figure \ref{subfig:std_scan_success_rates}, that when the standard deviations decrease to below 75\% of their nominal values, the predicted maximum pointwise success rate increases by a factor of two. Since the value of $H$, $k_{LH}$, and $T_i/T_e$ are the three main drivers of success, we test to see how adjusting their individual uncertainties contributes to the dramatic change in the maximum pointwise success rate when all uncertainties are reduced. Decreasing the standard deviation of the assumed $H$ distribution, without changing its mean, results in an improved predicted maximum pointwise success rate. As we assume the energy confinement time scaling law is more accurate, the best operating point moves to lower temperature, because the Q$>$2 surface is better defined, as can be seen in Figure \ref{subfig:H_std_ne_Te}. However, we see the opposite result when we change the standard deviation of $k_{LH}$, keeping the mean equal to one. Increasing our certainty about the L-H transition power threshold decreases our predicted maximum pointwise success rate because there are less Monte Carlo samples in the low temperature tail of the distribution. Correspondingly, increasing certainty about the L-H transition power threshold causes the ideal operating point to move to higher temperatures, where the L-H transition power threshold is more easily met, as shown in Figure \ref{subfig:kLH_std_ne_Te}. This suggests that low $k_{LH}$ values are highly favorable for H-mode access and the success conditions being met; we will explore this more in the following section. Decreasing the standard deviation of $T_i/T_e$, like confinement, has beneficial effect albeit much more slight. Such a decrease also favors slightly higher temperature solutions as can be seen in Figure \ref{subfig:temp_ratio_std_ne_Te}. Lower $T_i/T_e$ ratios reduce the success probability more than higher $T_i/T_e$ ratios increase it. Therefore, decreasing the standard deviation reduces the success rate.

\subsection{Effect of changing distribution means}\label{subsec:mean_scan}
\begin{figure}[htbp]
    \centering
    \begin{subfigure}{0.75\linewidth}
        \includegraphics[width=\linewidth]{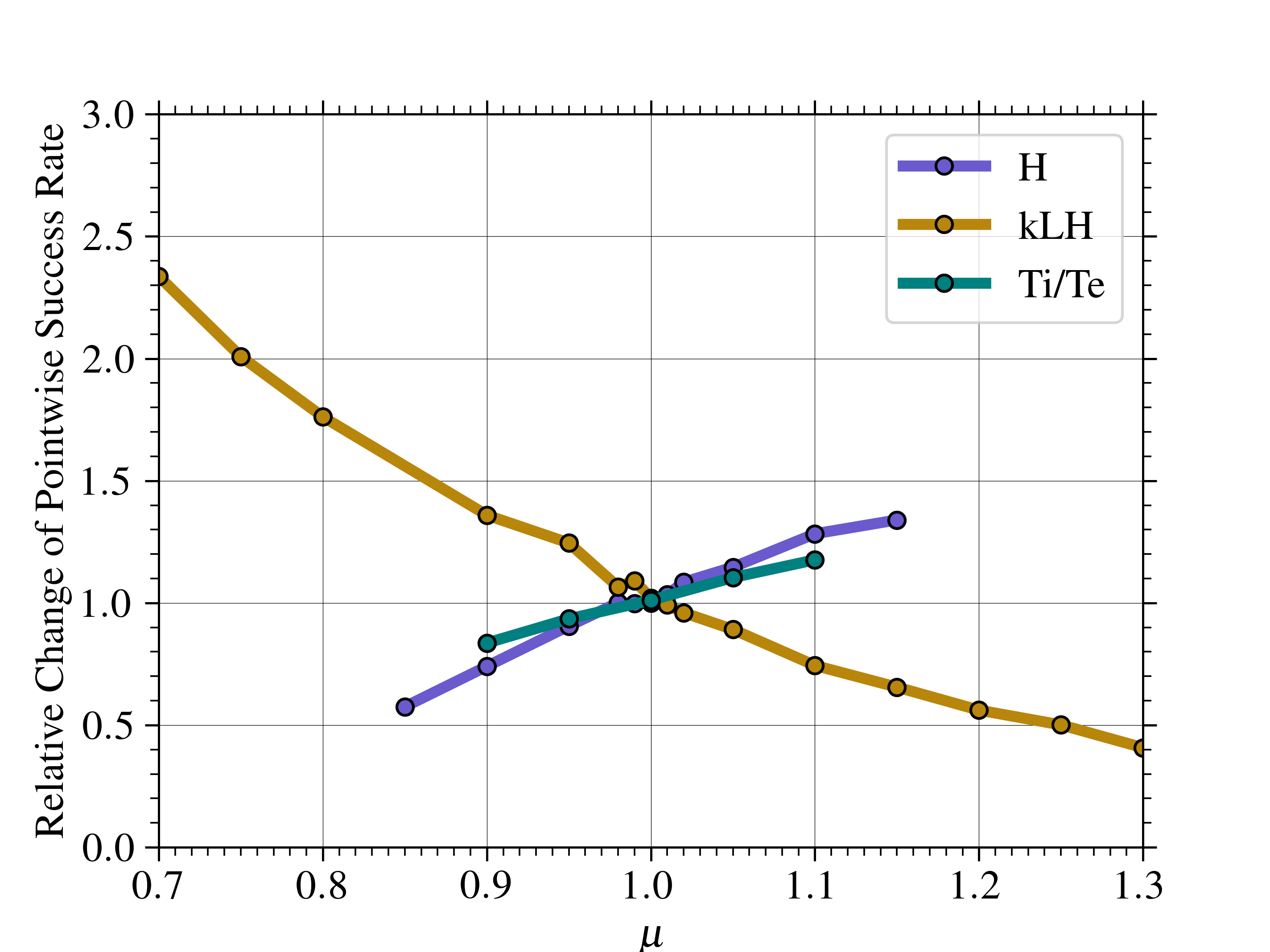}
        \caption{}
        \label{subfig:means_success_rates}
    \end{subfigure}
    \begin{subfigure}{0.32\linewidth}
        \includegraphics[width=\linewidth]{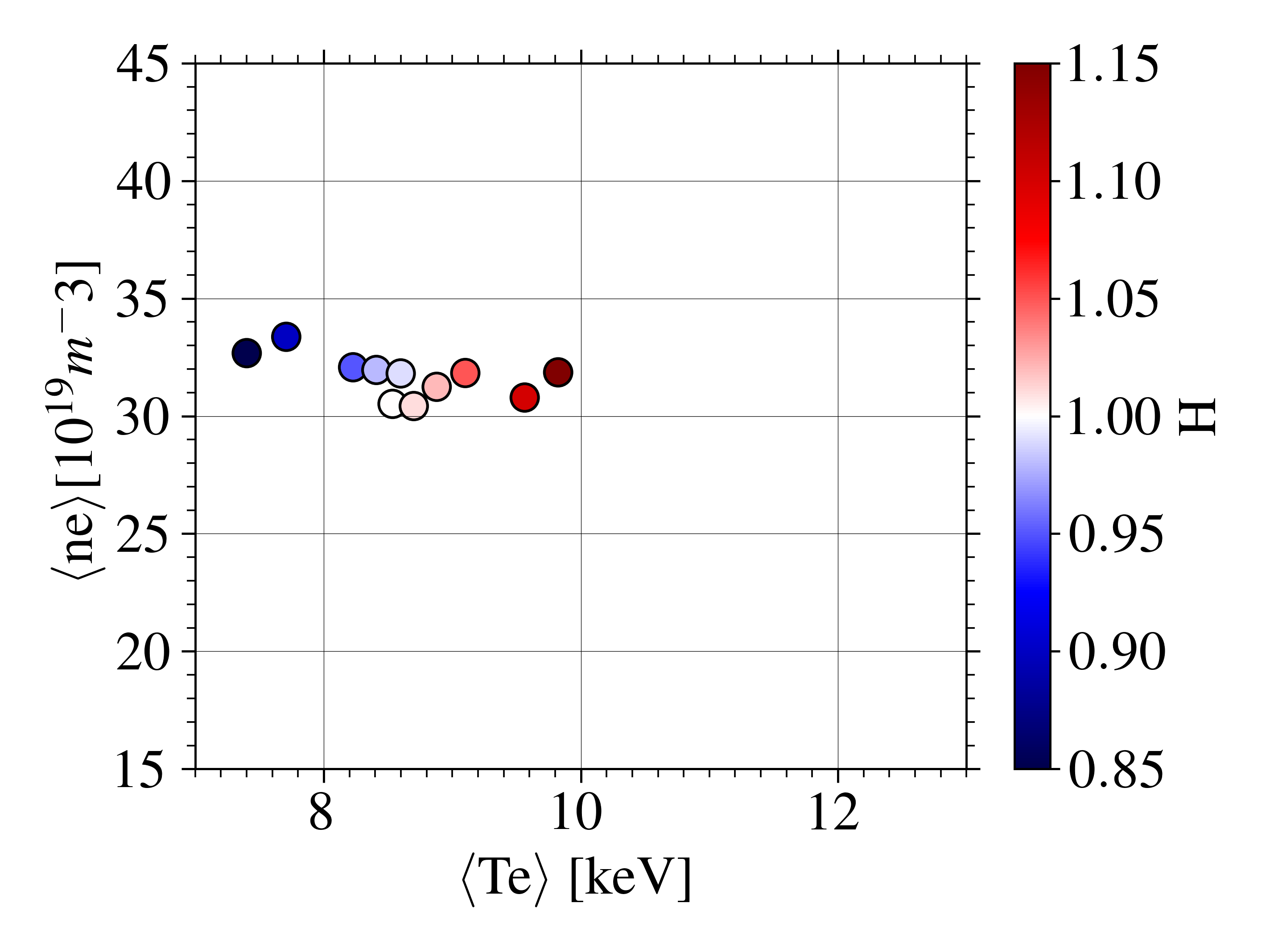}
        \caption{}
        \label{subfig:H_mean_ne_Te}
    \end{subfigure}
    \begin{subfigure}{0.32\linewidth}
        \centering
        \includegraphics[width=\linewidth]{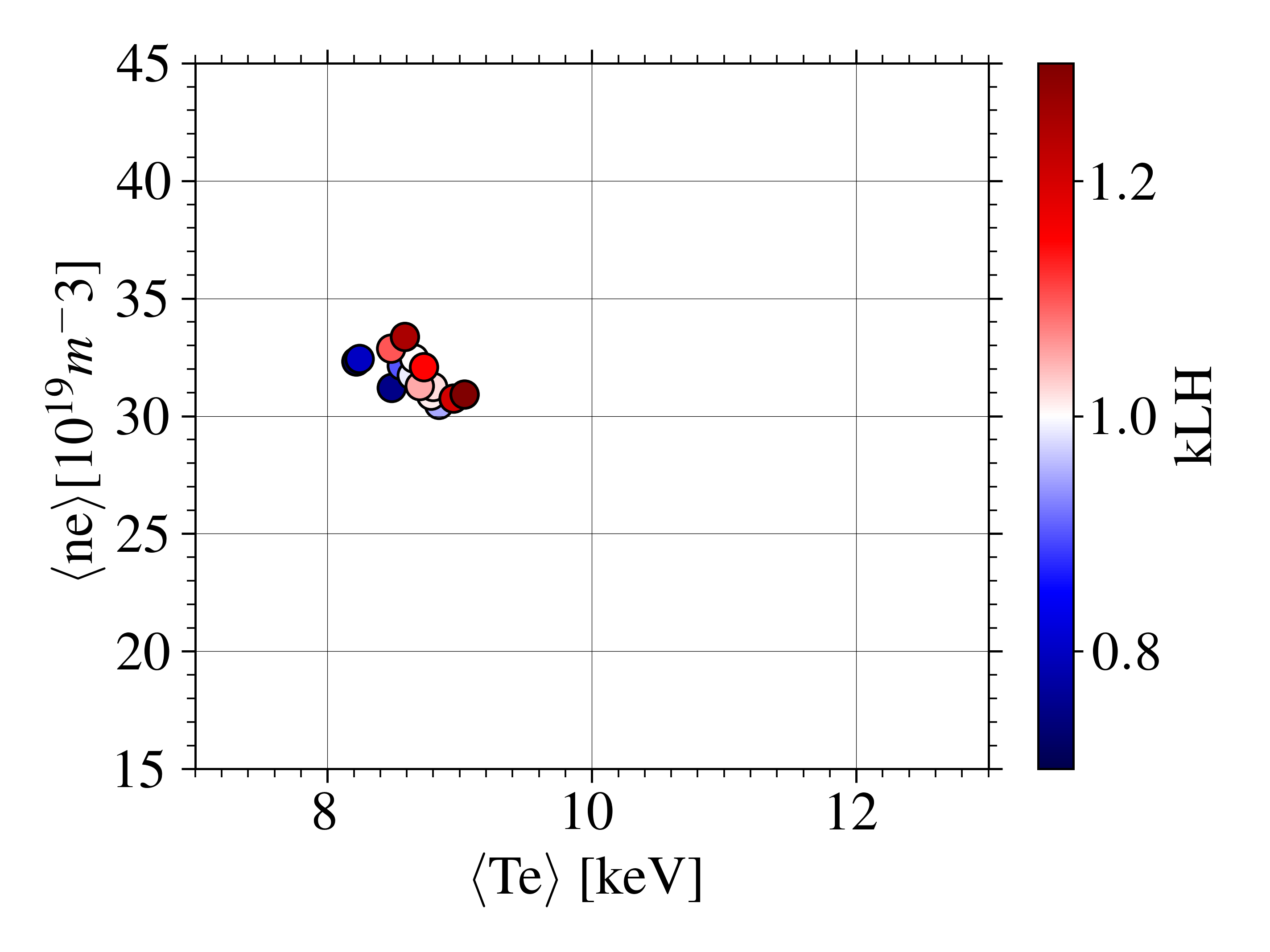}
        \caption{}
        \label{subfig:kLH_mean_ne_Te}
    \end{subfigure}
    \begin{subfigure}{0.32\linewidth}
        \centering
        \includegraphics[width=\linewidth]{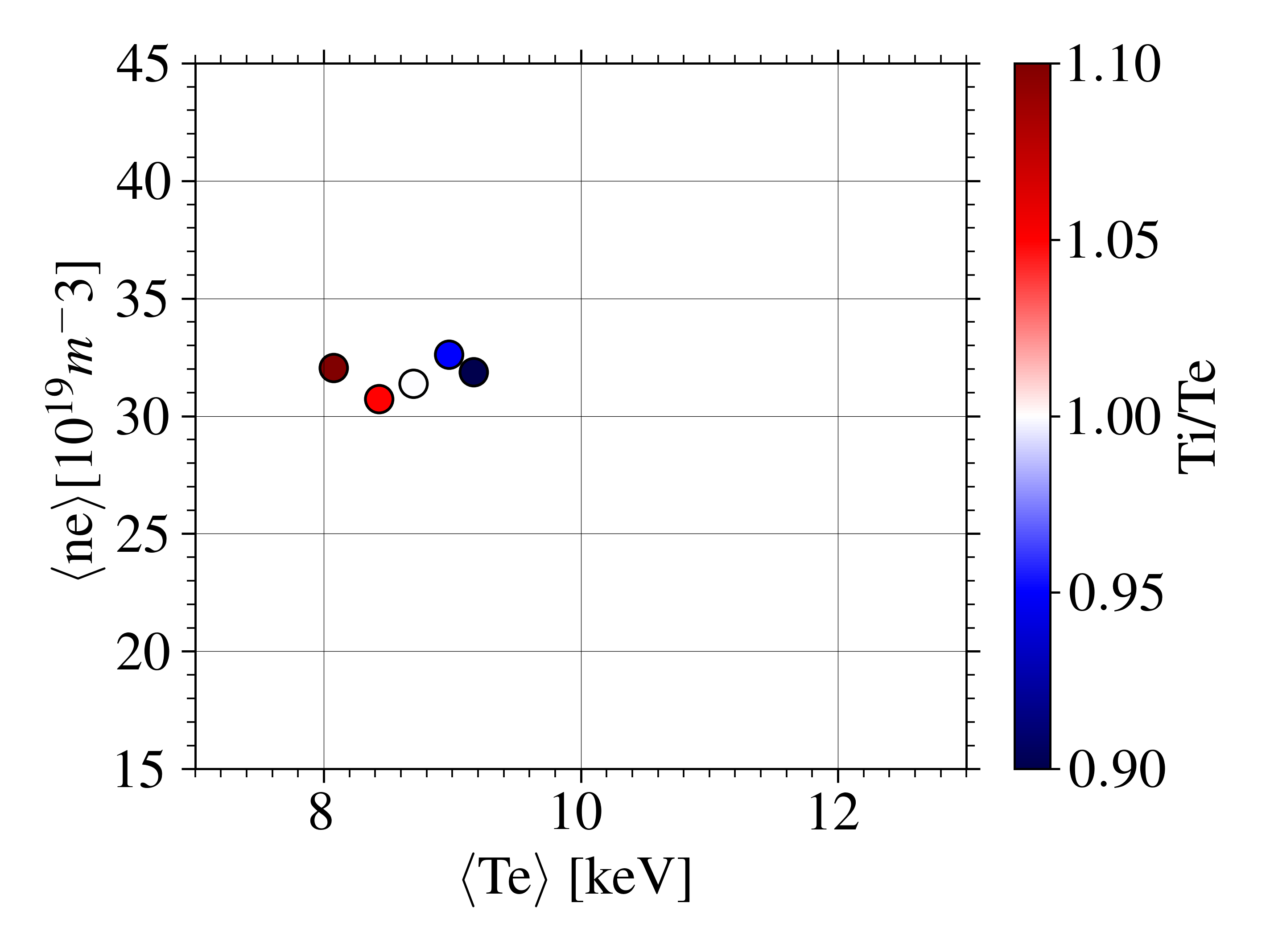}
        \caption{}
        \label{subfig:temp_ratio_mean_ne_Te}
    \end{subfigure}
    \caption{Effect of changing means of assumption distributions while preserving the default values of the relative standard deviations. Figure (a) looks at the change in the maximum pointwise success rate. Figures (b-d) show the location, in volume-averaged density - volume-averaged temperature space, of the maximum pointwise success rate. The maximum pointwise success rate increases as $H$ (blue) and $T_i/T_e$ (teal) increases and as $k_{LH}$ (orange) decreases. The optimal temperature increases with increasing $H$ and $k_{LH}$; it decreases with increasing $T_i/T_e$.}
    \label{fig:means_scan}
\end{figure}

As one might expect increasing the energy confinement time, decreasing the L-H transition power threshold, or increasing the ratio of $T_i/T_e$ causes the best pointwise success rate to increase. Figure \ref{subfig:means_success_rates} shows the effect of varying the mean of the distributions up and down by one standard deviation. The distributions retain the nominal relative standard deviations listed in Table \ref{table:assumptions}. This one sigma variation in the $k_{LH}$ mean results in a 100\% change in the maximum pointwise success rate, whereas the same degrees of variation in $H$ and $T_i/T_e$ result in only 50\% and 20\% changes respectively. Similar to the effect seen in Figure \ref{subfig:H_std_ne_Te}, an increase in the mean of $H$ results in a shift of the maximum pointwise success rate to a higher temperature both because it is easier to achieve the Q $>$ 2 success condition with better confinement and because the available 25 MW of auxiliary power can sustain a higher temperature.  Little variation of the operating point with the maximum pointwise success rate with changes in the $k_{LH}$ mean is seen in Figure \ref{subfig:kLH_mean_ne_Te}. Higher $T_i/T_e$ favors lower electron temperature solutions, as seen in Figure \ref{subfig:temp_ratio_mean_ne_Te}, to prevent higher stored energy than can be supported by 25 MW of auxiliary power

\subsection{Effect of edge impurity enrichment} \label{subsec:edge_enrichment}
\begin{figure}[ht]
    \centering
    \begin{subfigure}[b]{0.49\textwidth}
        \centering
        \includegraphics[width=\textwidth]{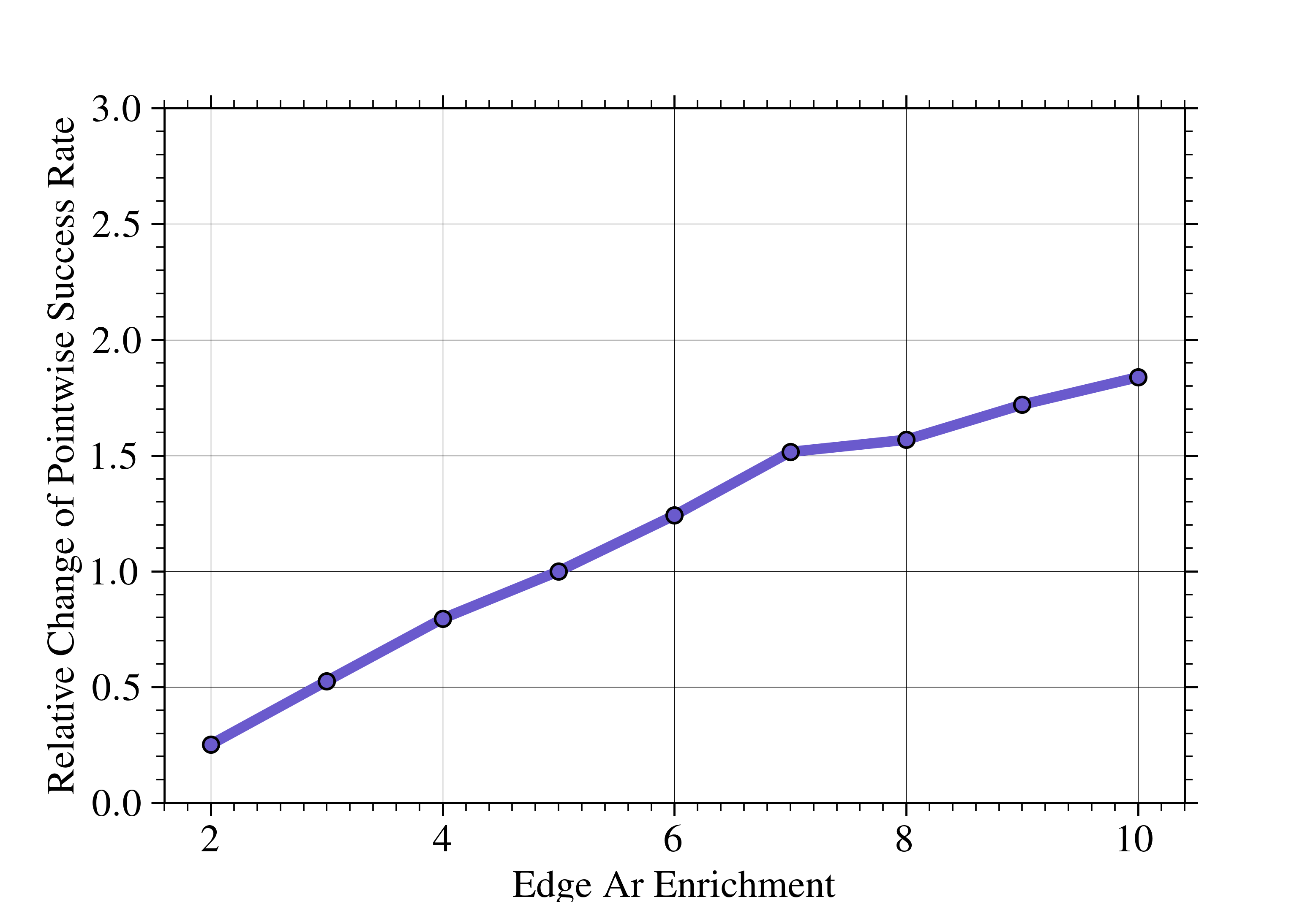} 
        \caption{}
        \label{subfig:Ar_enrichment_success}
    \end{subfigure}
    \hfill
    \begin{subfigure}[b]{0.49\textwidth}
        \centering
        \includegraphics[width=\textwidth]{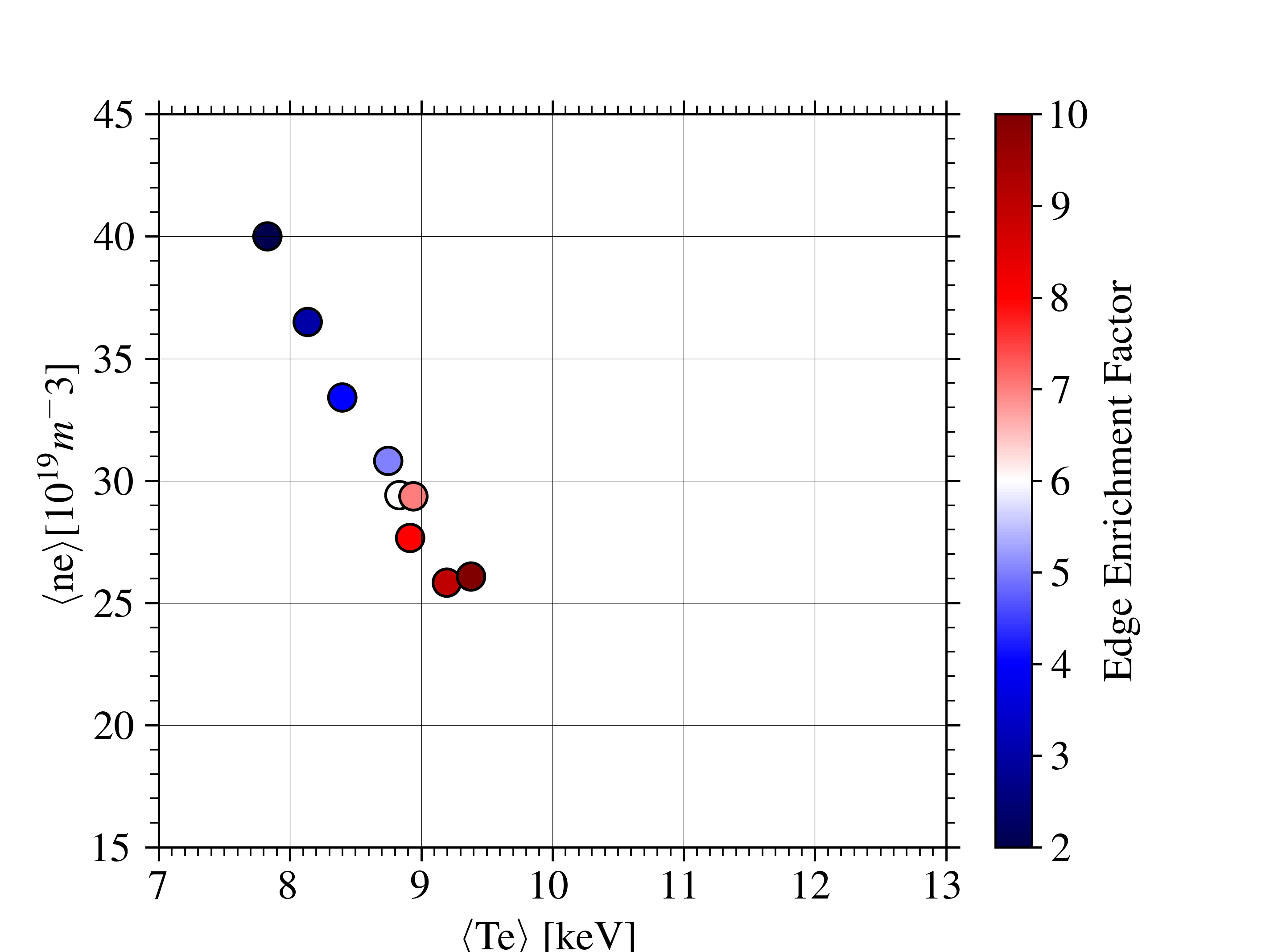}
        \caption{}
        \label{subfig:Ar_enrichment_ne_Te}
    \end{subfigure}
    \caption{Effect of the edge argon enrichment factor on (a) best pointwise success rate and (b) location in the volume-averaged density - volume-averaged temperature operating space of the best pointwise success rate. The edge argon enrichment is the concentration of argon available in the edge, used for radiating power away from the divertor, to the argon which dilutes the core. As less argon is transmitted from the edge to the core, the edge enrichment factor increases. Increasing the edge enrichment factor produces a notable improvement in performance and causes lower density higher temperature operating points to be more favorable. This is because increasing the edge enrichment factor makes finding a compatible core-edge solution easier.}
    \label{fig:Ar_enrichment}
\end{figure}

A compatible core-edge solution will be achieved in SPARC by injecting argon or neon impurities into the divertor region. In this work we consider the case of argon. Some fraction of the argon in the edge will travel through the last closed flux surface into the core where it increases dilution. As a reminder, the amount of argon is calculated to be consistent with the boundary solution, by using the Lengyel-Goedheer model \cite{stangeby_basic_2018} with a target electron temperature of 25 eV. The edge impurity enrichment is the ratio between the concentration of argon in the edge to the concentration in the core. Ideally, no argon from the edge would be transferred into the core. Therefore, the higher the edge impurity enrichment factor, the better the device is expect to perform. We scanned the range of enrichment values seen in ASDEX Upgrade H-modes as discussed in \cite{kallenbach_divertor_2024}. As shown in Figure \ref{subfig:Ar_enrichment_success}, increasing the argon enrichment by a factor of two also increases the maximum pointwise success rate by almost a factor of two. At higher argon enrichment values, a lower density higher temperature solution is favored, as can been seen in Figure \ref{subfig:Ar_enrichment_ne_Te}. Higher argon enrichment allows a larger fraction of the scrape-off-layer power to be radiated in the edge instead of directly conducted to the divertor and thereby increase its temperature. Higher edge densities also facilitate heat dissipation by radiation outside the last closed flux surface. Hence, lower enrichment factors require higher edge densities to radiate a sufficient fraction of the scrape-off-layer heat flux.

\subsection{Effect of engineering uncertainties}\label{subsec:engineering_uncertainties}

\begin{figure}[htbp]
    \centering
    \begin{subfigure}[b]{0.49\textwidth}
        \includegraphics[width=\textwidth]{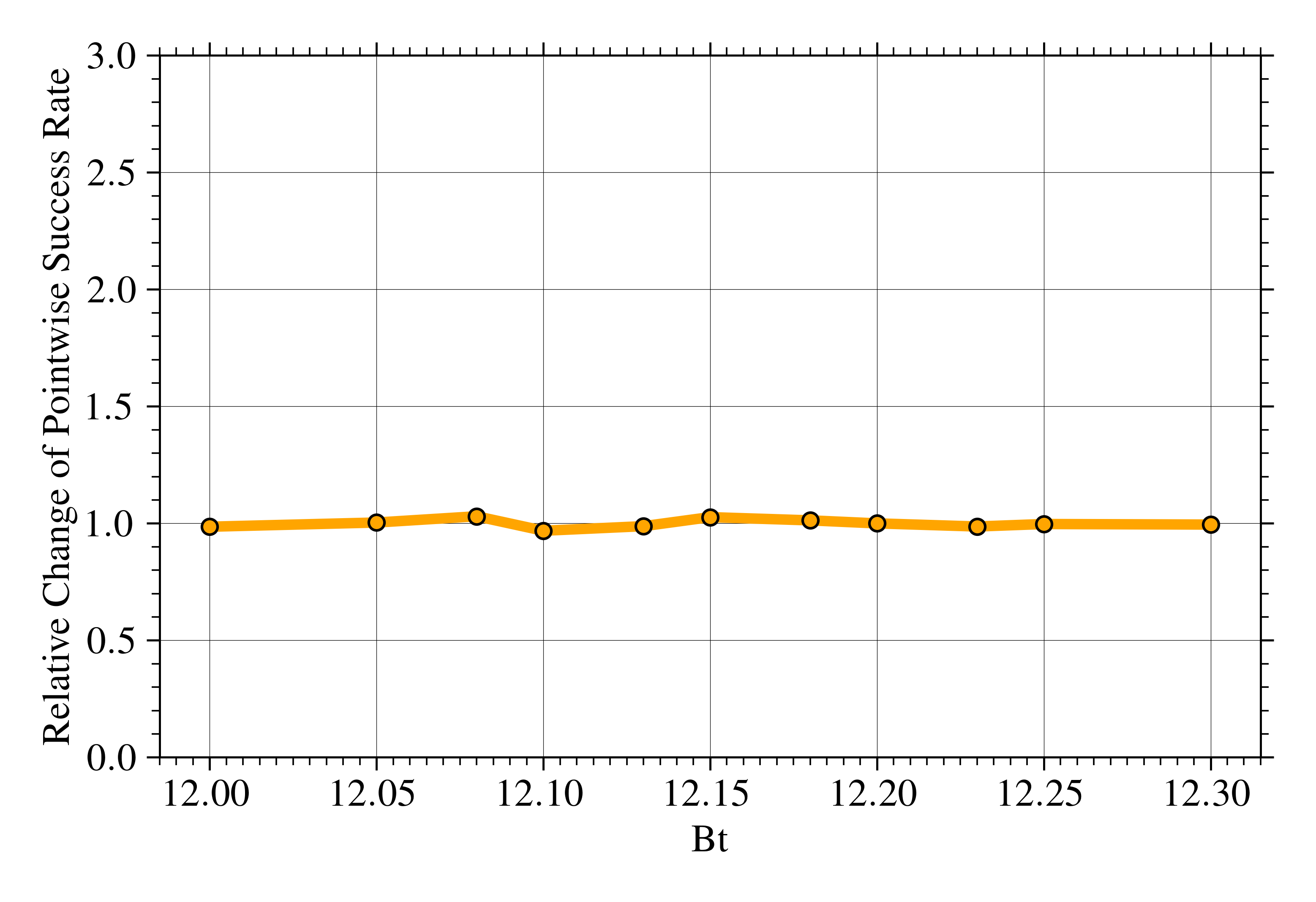} 
        \caption{}
        \label{subfig:Bt_success}
    \end{subfigure}
    \hfill
    \begin{subfigure}[b]{0.49\textwidth}
        \includegraphics[width=\textwidth]{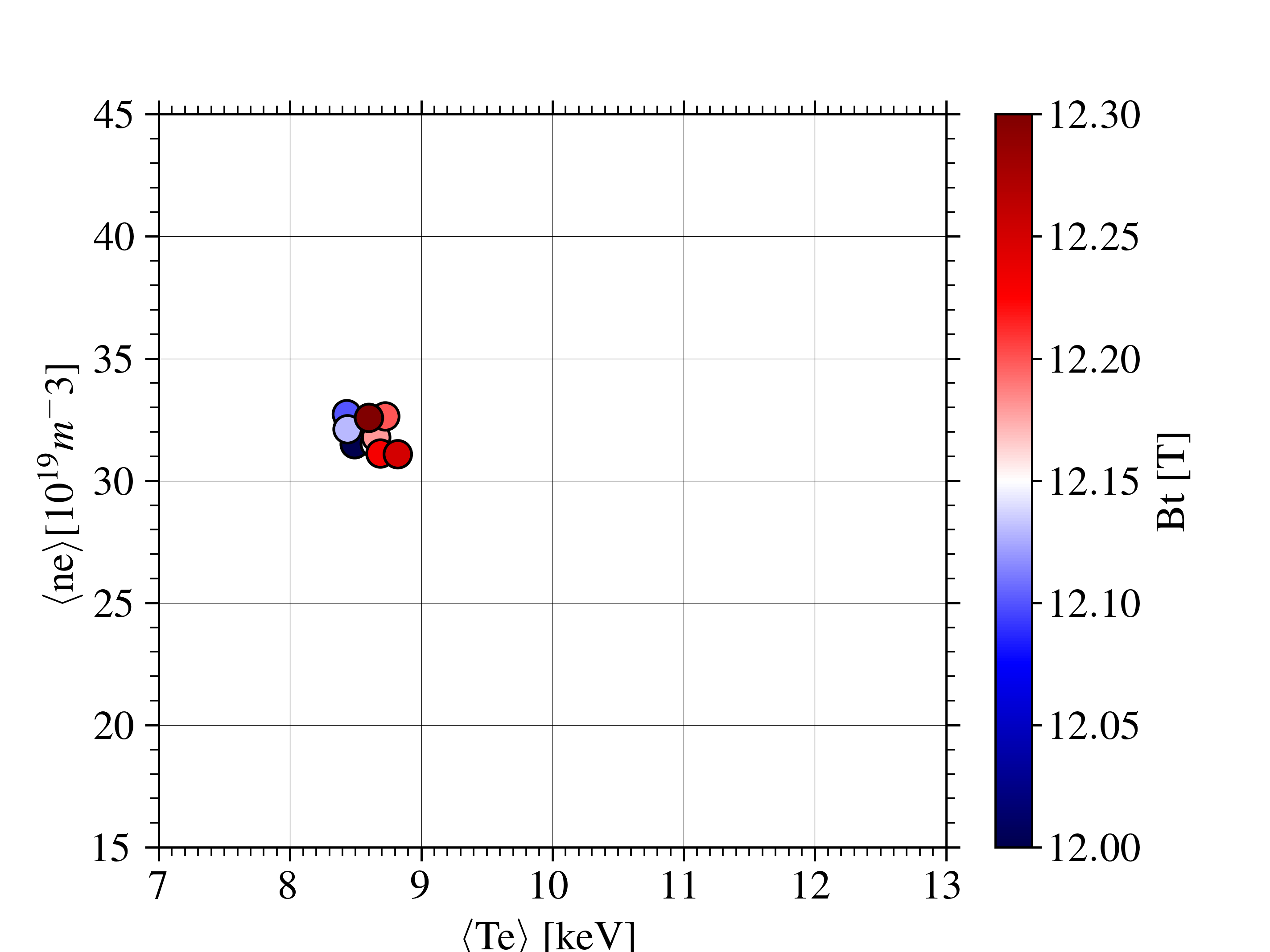}
        \caption{}
        \label{subfig:fixed_qstar_ne_Te.png}
    \end{subfigure}
    \caption{Effect of variations in the magnetic field strength, at fixed $q^*$, on (a) maximum pointwise success rates and (b) the location of the maximum pointwise success rate in volume-averaged density - volume-averaged temperature. Slight variations in magnetic field strength at fixed $q^\star$ have little impact on the ideal operation of SPARC.}
    \label{fig:Bt_scan}
\end{figure}

 In this section, we explore the impact of three potential variations in the machine's as-built design: the strength of the toroidal magnetic field, the  plasma current, and the available auxiliary power. Our interest in exploring lower plasma current is motivated by the opportunity to avoid disruptions. The expected maximum toroidal field strength, $B_0$, on SPARC is 12.2 T on axis \cite{creely_overview_2020}. Here, we explore the effect of small possible increases and decreases in the field strength while holding the kink safety factor, $q^* \propto \frac{B_0}{I_p}$, constant, where $I_p$ is the plasma current. For SPARC the kink safety factor is within 15\% of the safety factor at the 95\% flux surface. SPARC is designed for either 12.2 T full-field operation or 8 T reduced-field operation. These specific field strengths are required for coupling the ion cyclotron resonance heating (ICRH) antenna to the plasma effectively, using minority heating schemes with hydrogen (8 T) and helium-3 (12.2 T) \cite{lin_physics_2020}. Here we explore  slight changes on the 12.2 T case. We do not take into account a reduction in the efficiency of coupling RF power to the plasma that may be caused by these changes. We find that small, uniform perturbations in the magnetic field strength at fixed $q^*$ do not change the predicted performance, as can be seen in Figure \ref{subfig:Bt_success}, because the change in confinement is canceled out by the opposite change in ease of H-mode access. If plasma current is held constant, not $q^*$, the maximum pointwise success rate actually decreases slightly with increased magnetic field strength, since the predicted confinement time increases only weakly with magnetic field ($B^{0.15}$) but the L-H transition power threshold still increases notably ($B^{0.80})$.

\begin{figure}[ht]
    \centering
    \begin{subfigure}[b]{0.49\textwidth}
        \centering
        \includegraphics[width=\textwidth]{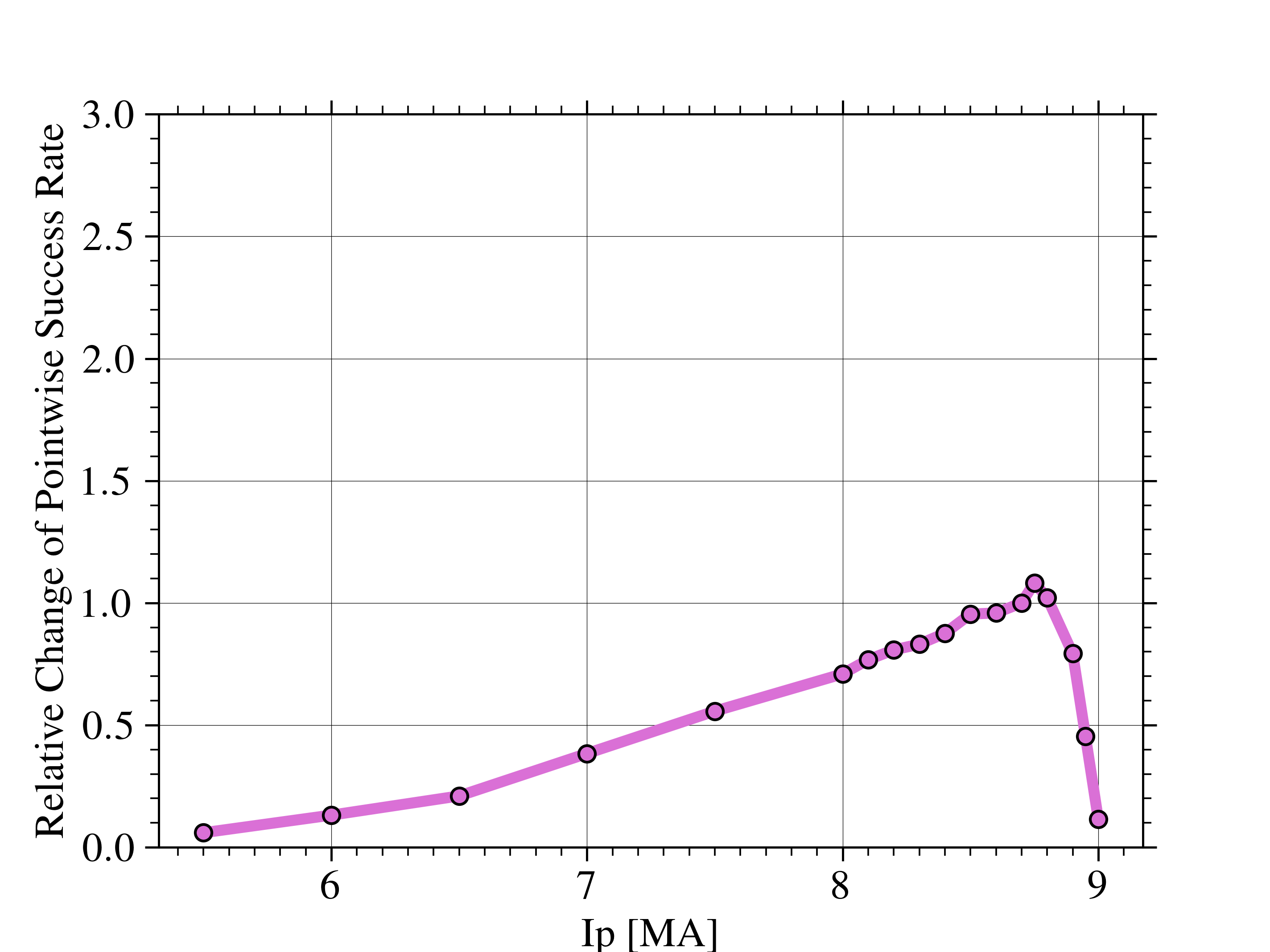} 
        \caption{}
        \label{subfig:Ip_success}
    \end{subfigure}
    \begin{subfigure}[b]{0.49\textwidth}
        \centering
        \includegraphics[width=\textwidth]{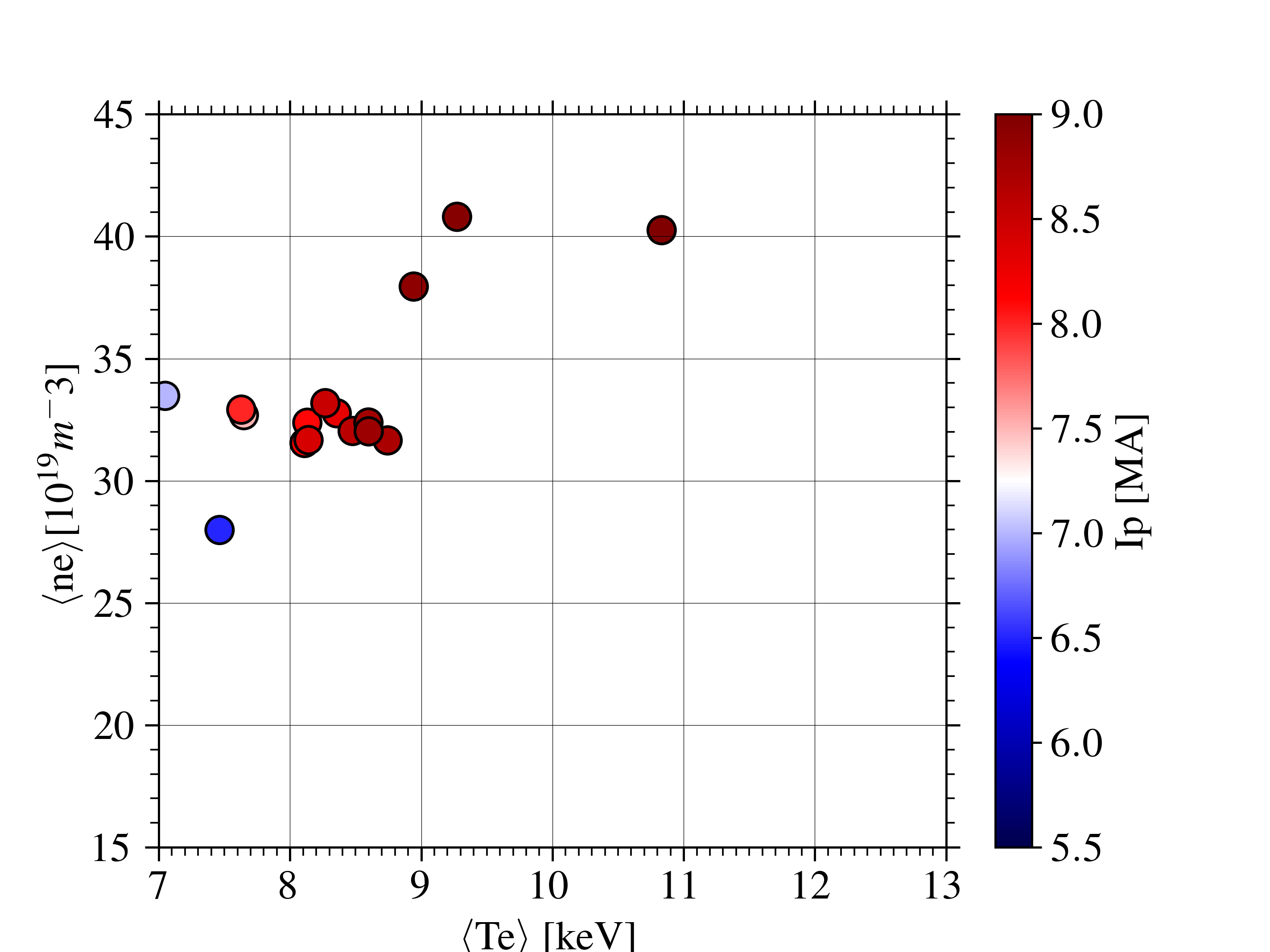}
        \caption{}
        \label{subfig:Ip_ne_Te}
    \end{subfigure}
    \caption{Effect of variations in the plasma current on (a) maximum pointwise success rates and (b) the location of the maximum pointwise success rate in volume-averaged density - volume-averaged temperature space. The SPARC PRD current is 8.7 MA. Lower current operation quickly degrades performance. Higher current operation is not possible because the central solenoid inductance is not sufficient to support a flattop period. }
    \label{fig:Ip_scan}
\end{figure}

Next, we consider only adjusting the plasma current, holding the magnetic field fixed at 12.2 T. SPARC has a planned plasma current of 8.7 MA for the PRD, a maximum performance, H-mode plasma. We scan a wide range of lower current options since operators can elect to run at lower than the maximum current. Figure \ref{subfig:Ip_success} shows that the maximum pointwise success rate improves linearly from the minimum current up to the planned 8.7 MA. As current decreases, operation moves to lower temperatures consistent with a reduction in the energy confinement time. This is the same trend as reducing the mean of H directly, as shown in Figure \ref{subfig:H_mean_ne_Te}. Higher current operation is unsuccessful in our model because the central solenoid does not have sufficient inductance to support a non-zero flattop time at higher current, resulting in a steep decrease of the maximum pointwise success rate for currents higher than the nominal PRD value of 8.7 MA. 

\begin{figure}[ht]
    \centering
    \begin{subfigure}[b]{0.49\textwidth}
        \centering
        \includegraphics[width=\textwidth]{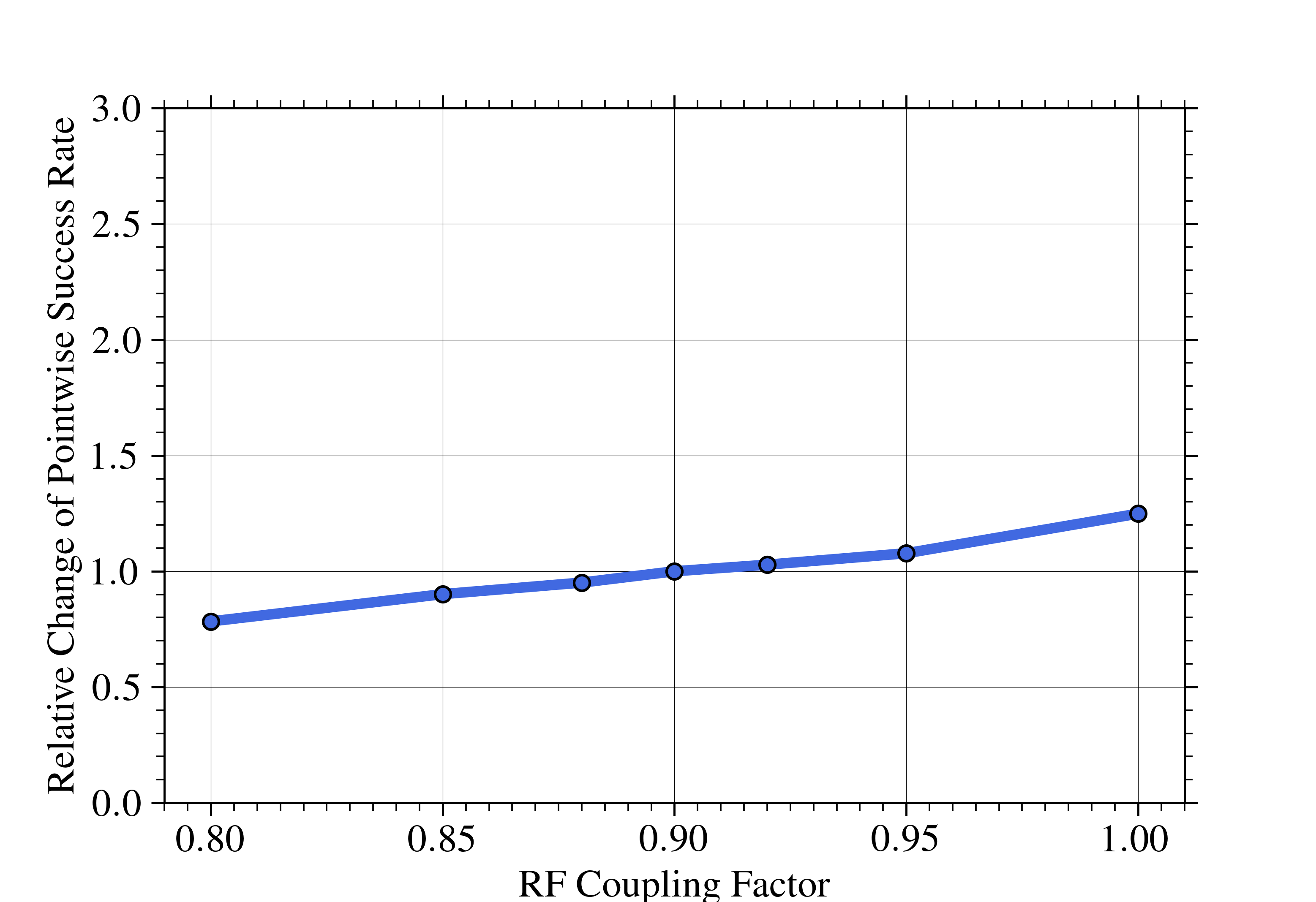} 
        \caption{}
        \label{subfig:rf_coupling_success}
    \end{subfigure}
    \hfill
    \begin{subfigure}[b]{0.49\textwidth}
        \centering
        \includegraphics[width=\textwidth]{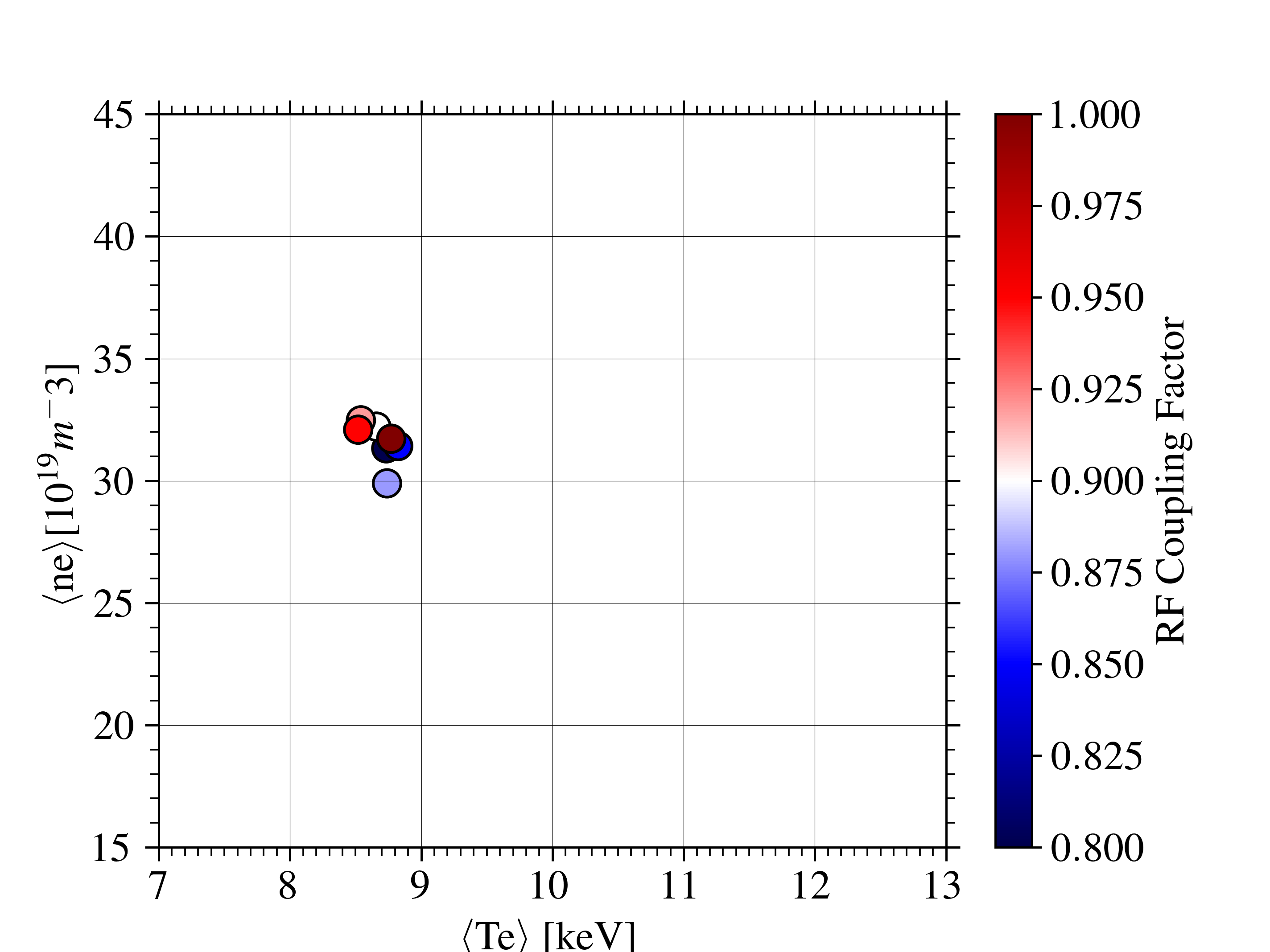}
        \caption{}
        \label{subfig:P_aux_ne_Te}
    \end{subfigure}
    \caption{The RF coupling factor is the fraction of launched RF power which is actually absorbed by the plasma. Nominally, we assume this value to be 90\%. Here, the effect of variations in the RF coupling factors is considered both for (a) the maximum pointwise success rates and (b) the location of the maximum pointwise success rate in volume-averaged density - volume-averaged temperature space.  Small perturbations in this factor suggest only slight changes the performance and best operating point location.}
    \label{fig:rf_coupling_scan}
\end{figure}

There are two possible ways in which the auxiliary power could be different than expected. Either the available auxiliary power to be launched could change or the fraction of the launched power which is absorbed by the plasma could change. First, we look at  the case of a smaller fraction of the launched power being coupled. As a rough approximation, it is generally assumed for the SPARC PRD that 90\% of the launched RF power is coupled to the plasma \cite{body_sparc_2023}. In Figure \ref{fig:rf_coupling_scan}, the fraction of launched power coupled to the plasma is scanned from 80\% to 100\%. The maximum pointwise success rate linearly increases with the coupling fraction. It is important to note that, throughout the whole range of this scan, the coupled auxiliary power only changes by 2.5 MW. The operating point with the maximum pointwise success rate does not substantially shift as higher stored energy operating points require strong increases in the amount of input power to become accessible because of the power degradation term in the energy confinement scaling laws.

\begin{figure}[ht]
    \centering
    \begin{subfigure}[b]{0.49\textwidth}
        \centering
        \includegraphics[width=\textwidth]{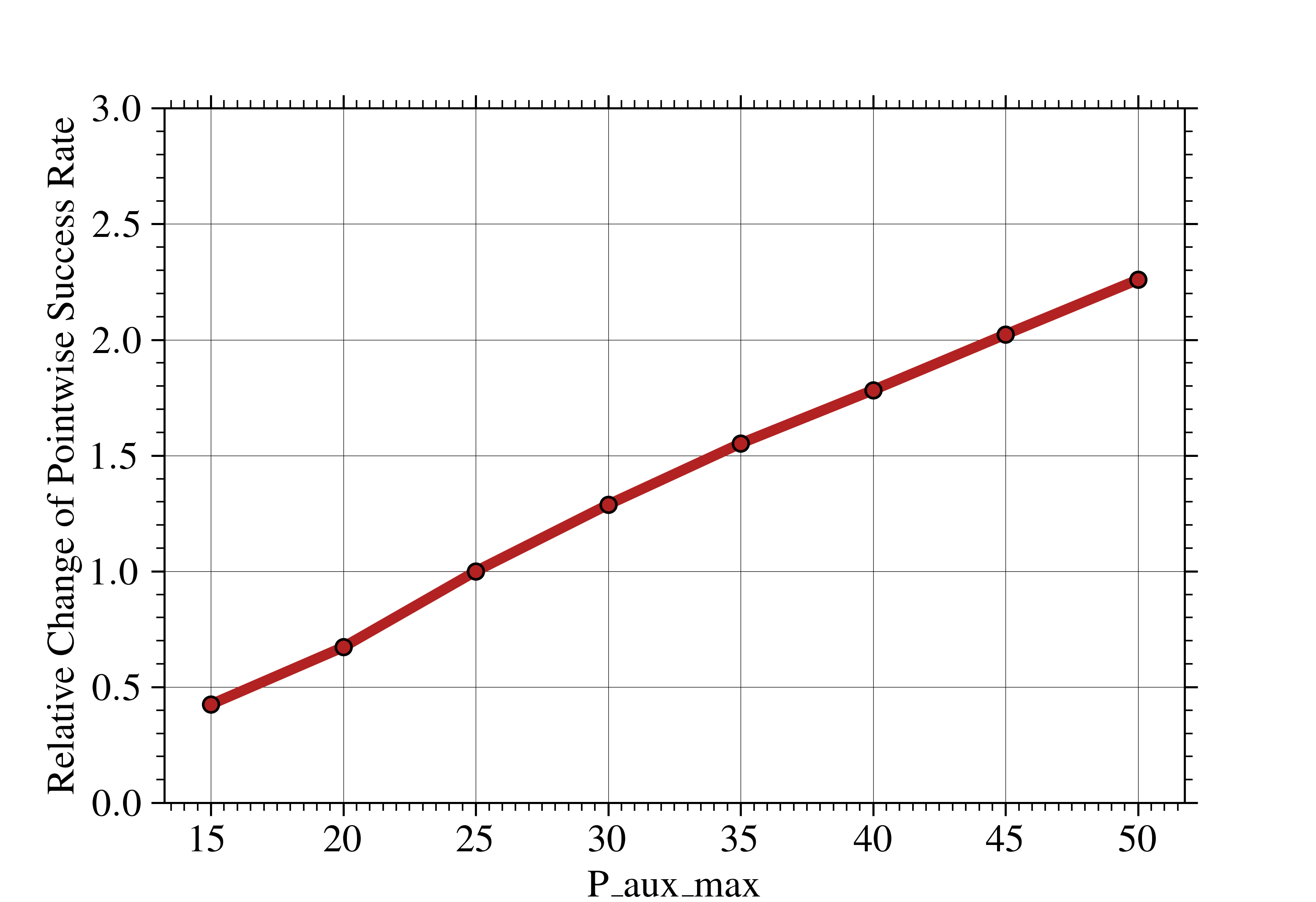} 
        \caption{}
        \label{subfig:P_aux_success}
    \end{subfigure}
    \hfill
    \begin{subfigure}[b]{0.49\textwidth}
        \centering
        \includegraphics[width=\textwidth]{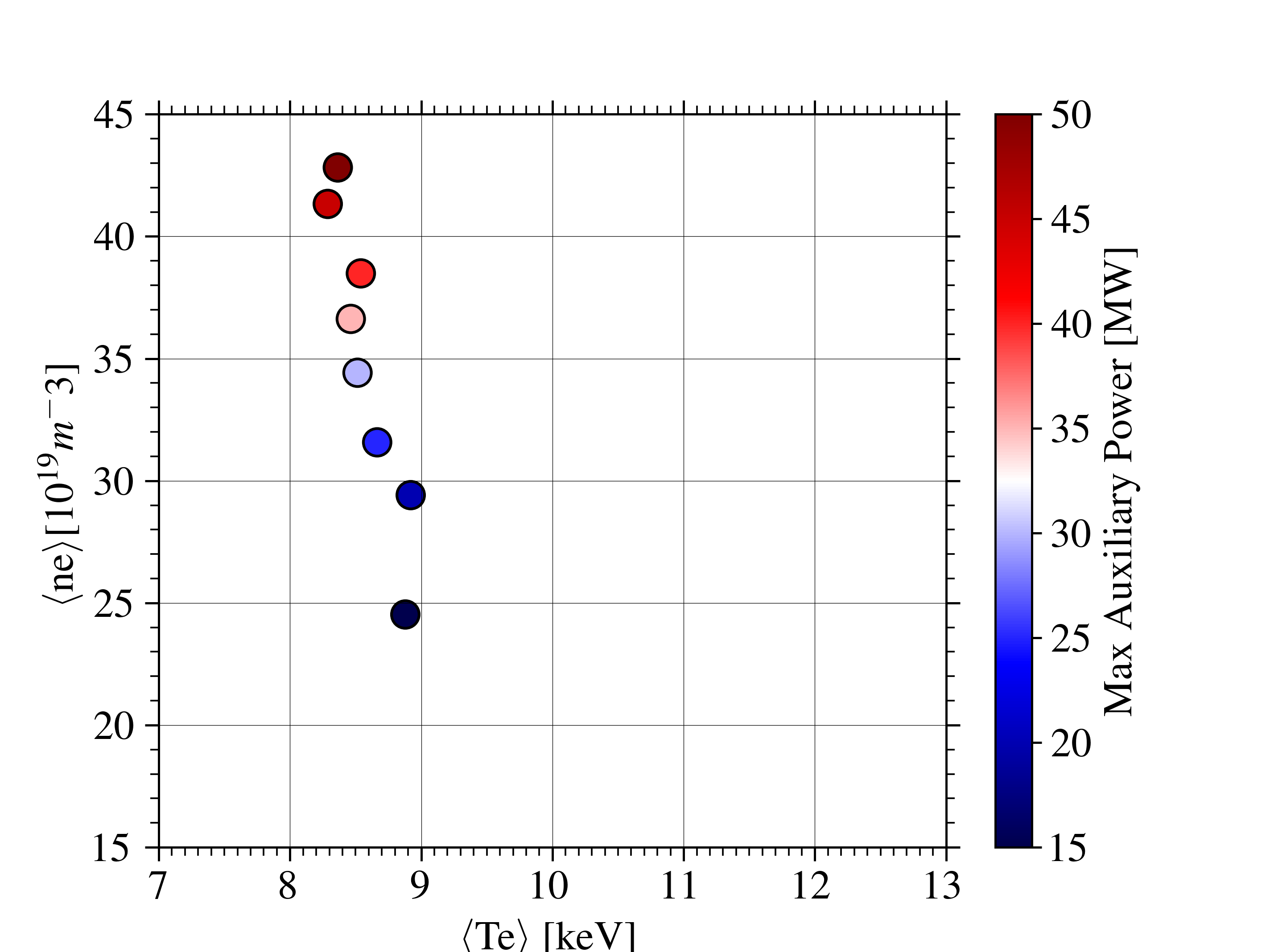}
        \caption{}
        \label{subfig:P_aux_ne_Te}
    \end{subfigure}
    \caption{Effect of variations in the launched auxiliary power on (a) maximum pointwise success rates and (b) the location of the maximum pointwise success rate in volume-averaged density - volume-averaged temperature space. Increasing the available auxiliary power launched up to and above the planned 25 MW has a strong predicted performance improvement due at least in part to the importance of being able to access H-mode. The ideal operating point density increases with increasing available power as higher plasma stored energies are achievable.}
    \label{fig:P_aux_scan}
\end{figure}

The second way to change the amount of auxiliary power available is to change how much is launched. We look at the results of changing the amount of auxiliary power launched within the range of 15MW to 50MW, displayed in Figure \ref{fig:P_aux_scan}. It should be noted that the 50 MW upper bound is not feasible and should be thought of as a numerical exercise. A reduction from 25 MW of available launched auxiliary power to 15 MW results in a halving of the maximum pointwise success rate. The maximum pointwise success rate increases approximately linearly with increasing available launched power.  As more power becomes available, the best operating point increases in volume-averaged density. Recall that the necessary input power is calculated by dividing the plasma stored energy by the energy confinement time. Thus more available power means that for a given energy confinement time, a higher plasma stored energy can be supported. The power degradation term in the energy confinement scaling likely explains why there is not also a corresponding increase in the ideal volume-averaged temperature. The sensitivity to both changing the fraction of launched auxiliary power coupled to the plasma as well as changing the amount of auxiliary power that is launched display SPARC's sensitivity to available auxiliary power.

\section{Discussion} \label{sec:Discussion}
In this work, we strive to maximize SPARC's rate of complying with its goals in a 8.7MA DT H-mode scenario. We performed a Monte Carlo analysis varying ten parameters central to POPCON modeling predictions to see how such variations impacted the maximum pointwise success rate. Of these, two had particularly dramatic effects: $H$, the energy confinement time factor, and $k_{LH}$, the confinement threshold factor. Once uncertainties are taken into account, the location of the high pointwise success rate in volume-averaged density - volume-averaged temperature space is a broad maximum. This area balances having H-mode access, having a sufficient confinement time to support the plasma stored energy with available input power, and limiting dilution from argon injected to ensure a robust boundary solution. Our findings can be summarized succinctly as being that future devices should not rely on operating close to boundaries (e.g. the H-mode transition) which have substantial uncertainty. Equivalent deterministic POPCON analysis gives a different ideal operating point than a statistical one.

We find the success rate increases dramatically over the maximum pointwise success rate if multiple shots are available to explore the volume-averaged density - volume-averaged temperature space. This emphasizes testing is key to achieving reliable performance. One key aspect of future work would be developing a tool that updates the predictions of the pointwise success rate in real time as information comes in from previous shots. This could include corrections to the energy confinement time predictions and the H-mode transition power threshold via databases of SPARC data.

Bayesian optimization provides an efficient way to find the operating point with the maximum pointwise success rate, in comparison to a brute force search or use of the Powell algorithm. This allowed us to perform detailed scans varying the input assumptions as well as other engineering parameters. An even greater speed up can be achieved by using a basic multi-fidelity approach. Here, the surrogates are first trained on pointwise evaluations with few Monte Carlo samples, which results in large uncertainties on each evaluation. The solution is then refined by sampling with increasing number of Monte Carlo samples. Thus, efficient solutions can be found relatively quickly, without the overshooting of the maximum pointwise success rate due to statistical fluctuations that characterize lower fidelity evaluations.

The predicted pointwise success rate at any given volume-averaged density - volume-averaged temperature operating point increases dramatically if the L-H transition threshold power can be decreased. Possible routes to systematically reducing the L-H transition power threshold include adjusting the divertor geometry and X-point location \cite{gohil_lh_2011, ma_h-mode_2012, maggi_lh_2014}, decreasing the effective impurity charge (Zeff) \cite{takizuka_roles_2004, bourdelle_l_2014}, increasing the fraction of heat exhaust that is through the ion channel \cite{ryter_experimental_2014, schmidtmayr_investigation_2018}, increasing the fraction of the fuel that is tritium \cite{righi_isotope_1999}, using a double null divertor configuration \cite{meyer_h-mode_2005}, adjusting the plasma upper and lower triangularity \cite{maggi_lh_2014}, or taking advantage of hysteresis effects before the H-L back transition through trajectory optimization \cite{chatthong_understanding_2016}. For a more thorough discussion of these variables not captured in the H-mode power threshold scaling law model used here, please see \cite{hughes_projections_2020}. While reducing uncertainties on the L-H transition power threshold is not enough to improve the maximum pointwise success rate, such a reduction on the energy confinement time uncertainty, around the value nominally predicted by the scaling law, would be very beneficial. It may also be possible to take advantage of the large, natural shot-to-shot fluctuations that mean different Monte Carlo ``samples" are reflective of different shots. Predictions made with physics-based core turbulent transport models offer the opportunity of improving our certainty for a given planned discharge.

Another key consideration to ensuring SPARC's success is finding a divertor boundary solution compatible with a successful core one. The ability to reduce the amount of argon, which is pumped into the divertor, that enters the core would be enormously beneficial in this regard. Engineering uncertainty analyses shows the importance of having the full 25 MW of launched auxiliary power installed on the device by the time these key shots are attempted. Potentially increasing the auxiliary power could help if unexpected challenges are encountered using 25 MW; high power solutions would operate at higher densities. Unfortunately, reducing current to mitigate disruptions rapidly degrades the predicted device performance.

One limitation of the work as completed is that empirical models do not fully capture core transport behavior. Because of critical gradient behavior, at higher temperatures the power degradation term of the energy confinement scaling law becomes stronger than current laws predict. In fact, some temperatures shown in statistical POPCONs may not be achievable at all. Physics-based modeling has not found a consistent solution that produces temperatures in excess of 11 keV \cite{muraca_integrated_2025}. An additional limitation is that temperature is not an engineering parameter. Instead, an operator has direct control over the input auxiliary power. POPCONs in density - auxiliary power space are a promising direction for future research. In this work, we assume that density can be treated as an engineering parameter. However, it may be difficult to fuel such that density dramatically increases once in H-mode operation \cite{ma_h-mode_2012}. Work is currently ongoing looking into trajectory optimization ensuring that the desired steady-state operating conditions can be achieved. It should be noted that pointwise success rate is not directly proportional to the overall success rate since the size of the successful area can increase and decrease. While expensive, further analysis of how the total success rate changes with changing assumptions could be very fruitful.

Other valuable future work includes considering the consistency of the boundary solution of assumed profiles with EPED \cite{snyder_first-principles_2011} or an EPED neural network \cite{meneghini_self-consistent_2017, hall_multi-fidelity_2024}. Improvements to the L-H transition power threshold model, including a metal-wall scaling currently under development, could have large, likely positive, impacts on the results. In addition to varying a scalar prefactor on the L-H transition power threshold, another prefactor could be put on the location of the density minimum which also has substantial uncertainty \cite{ryter_experimental_2014}. Finally, similar statistical POPCON studies could be carried out with the goal of optimizing other operating regimes in SPARC or other earlier design stage machines.

\section{Acknowledgments}
The authors would like to acknowledge helpful conversations within members of the Performance and Transport MIT-CFS RPP group, particularly Jerry Hughes, Andr\'es Miller, and Conor Perks. They also would like to thank Dennis Whyte for his suggestion of exploring engineering parameter uncertainties. ChatGPT4 and CoPilot were used in code generation, formatting, and word choice enhancement. This work is supported by the National Science Foundation Graduate Research Fellowship under Grant No. 2141064 and by Commonwealth Fusion Systems, under RPP020. 

\printbibliography
\end{document}